\documentclass[trackchanges, twocolumn]{aastex7}

\usepackage{anyfontsize}
\usepackage{xcolor}
\usepackage{newtxmath}
\usepackage{subcaption}

\shorttitle{Sudden Change in the Global Magnetic Field of YZ Cet}
\shortauthors{Biswas et al.}

\begin{document}

\title{Discovery of a rapidly evolving global magnetic field in the M-dwarf YZ Cet \\and constraints on the magnetic field of its planet YZ Cet b}

\author[0000-0002-1741-6286]{Ayan Biswas}
\altaffiliation{Based on observations obtained at the Canada-France-Hawaii Telescope (CFHT) which is operated by the National Research Council (NRC) of Canada, the Institut National des Sciences de l'Univers of the Centre National de la Recherche Scientifique (CNRS) of France, and the University of Hawaii. The observations at the CFHT were performed with care and respect from the summit of Maunakea which is a significant cultural and historic site.}
\affiliation{Department of Physics, Engineering Physics \& Astronomy, Queen’s University, Kingston, Ontario K7L 3N6, Canada}
\affiliation{Department of Physics \& Space Science, Royal Military College of Canada, PO Box 17000, Station Forces, Kingston, ON K7K 7B4, Canada}
\affiliation{National Centre for Radio Astrophysics, Tata Institute of Fundamental Research, Ganeshkhind, Pune-411007, India}
\email[show]{ayan.biswas@queensu.ca}

\author[0000-0002-9023-7890]{Colin P. Folsom}
\affiliation{Tartu Observatory, University of Tartu, Observatooriumi 1,Toravere, 61602, Estonia}
\email{folsomcp@gmail.com}

\author[0000-0003-2088-0706]{James A. Barron}
\affiliation{Department of Physics, Engineering Physics \& Astronomy, Queen’s University, Kingston, Ontario K7L 3N6, Canada}
\email{j.barron@queensu.ca}

\author[0000-0002-1854-0131]{Gregg A. Wade}
\affiliation{Department of Physics \& Space Science, Royal Military College of Canada, PO Box 17000, Station Forces, Kingston, ON K7K 7B4, Canada}
\affiliation{Department of Physics, Engineering Physics \& Astronomy, Queen’s University, Kingston, Ontario K7L 3N6, Canada}
\email{Gregg.Wade@rmc.ca}

\author[0000-0002-2558-6920]{Stefano Bellotti}
\affiliation{Leiden Observatory, Leiden University, PO Box 9513, 2300 RA, Leiden, The Netherlands}
\affiliation{Institut de Recherche en Astrophysique et Plan\'etologie, Universit\'e de Toulouse, CNRS, IRAP/UMR 5277, 14 avenue Edouard Belin, F-31400, Toulouse, France}
\email{bellotti@strw.leidenuniv.nl}


\author[0000-0002-1216-7831]{Corrado Trigilio}
\affiliation{Osservatorio Astrofisico di Catania, INAF, Via S. Sofia 78, I-95123 Catania, Italy}
\email{corrado.trigilio@inaf.it}

\correspondingauthor{Ayan Biswas}


\begin{abstract}

We present a spectropolarimetric study of the nearby M4.5V exoplanet host star YZ~Cet, based on near-infrared observations obtained with the SpectroPolarimètre InfraRouge (SPIRou) at the Canada–France–Hawaii Telescope. We detect striking changes in the large-scale magnetic field strength and geometry over the course of just a few stellar rotations, a level of short-term global magnetic field evolution rarely reported in M dwarfs. We modeled the temporal variation of the longitudinal magnetic field using the Gaussian regression process, which allowed us to robustly determine the stellar rotation period and quantify the evolution timescale of the magnetic field. Independent Zeeman Doppler Imaging reconstructions of the two epochs confirm a significant reconfiguration of the star’s global magnetic strength and topology. The detection of a weaker complex axisymmetric magnetic field (Mean $|B| \sim 201$ G), which changes into stronger non-axisymmetric dipole dominated field (Mean $|B| \sim 276$ G) over a few rotation cycles, is in contrast to results from similar fully convective M-dwarf stars. YZ~Cet is known to exhibit polarized radio bursts potentially driven by auroral radio emission from star-planet interaction (SPI). By combining our magnetic maps with recent radio observations, we refine the constraints on the magnetic field strength of the innermost planet, YZ~Cet~b. These results underscore the importance of monitoring stellar magnetic variability to interpret multi-wavelength SPI signatures, and to characterize the magnetospheres of potentially habitable exoplanets.

\end{abstract}



\section{Introduction}

Magnetic fields play a crucial role for both the interior and atmospheric properties of M dwarfs, the most common stars found in the galaxy \citep{Winters2019, Kochukhov2021}. These magnetic fields affect not only the star (e.g. \citealt{Donati2009, Reiners2012, Kochukhov2021, Reiners2022}), but also any associated exoplanets by potentially causing a star-planet interaction (SPI), affecting their orbital parameters, or impacting the habitability of nearby planets \citep{Vidotto2014, Strugarek2015, Hebrard2016}. Numerous planetary systems have been detected around M dwarfs in the last two decades, making it increasingly necessary to monitor the magnetic fields of these stars \citep{Bonfils2013, Gaidos2016}. Exoplanetary magnetic fields are believed to play an important role in the habitability of such planets, in terms of protection against stellar radiation and the evolution of atmospheres \citep{Airapetian2017}. 

In the solar system, magnetized planets capture charged particles from the solar wind, producing aurorae visible in various wavebands. These planets emit strongly polarized radio bursts linked to auroras, generated by Electron Cyclotron Maser Emission (ECME), known as Auroral Radio Emission (ARE, e.g. \citealt{Mutel2008}). Aurorae are key indicators of SPI in magnetized planets and exoplanets. Detection of SPI can give an indirect estimate of planetary magnetic fields, given a reliable model of the stellar magnetic field \citep{Trigilio2023, Pineda2023}. Thus, it is crucial to monitor the long-term variability of magnetic fields of exoplanet-hosting systems to reliably model the planetary magnetospheres based on multiwavelength observations. 

For stars with non-aligned magnetic and rotation axes, the longitudinal magnetic field $B_{\ell}$ is expected to show rotational modulation \citep{Donati2009}. For weakly active M dwarfs with long rotation periods, the variation of $B_{\ell}$ observed from spectropolarimetric studies serves as an essential tool to determine the stellar rotation period \citep{Fouque2023, Donati2023}. The rotational modulation of the intensity and polarization spectra of M dwarfs can be exploited to reconstruct vector maps of the surface magnetic field using a tomographic technique such as Zeeman Doppler Imaging (ZDI; \citealt{Donati2006, Kochukhov2021}). Although the $B_{\ell}$ measurements of M dwarfs are observed to evolve on a short timescale of few rotation cycles (e.g. \citealt{Donati2023, Fouque2023}), repeated ZDI studies based on observations obtained 1--2 years apart revealed no major systematic evolution of global magnetic fields in general \citep{Kochukhov2021, Bellotti2024}. { However, global magnetic field evolution is observed in several early-type M-dwarfs, for example in the well-known M dwarf star AD Leo}, where a long-term evolution of the global magnetic field is observed, which features a decrease in axisymmetry and is accompanied by a weakening of the longitudinal field \citep{Lavail2018, Bellotti2023}. 

ZDI mapping has only recently been performed on a small number of slowly rotating ($P_{\mathrm{rot}}\approx40-180$\,d) fully convective (FC) M dwarfs. The mapped stars include Proxima Centauri \citep{Klein2021} and five stars analyzed by \cite{Lehmann2024} from the SPIROU Legacy Survey \citep{Donati2020}. The ZDI maps show that slowly rotating FC stars possess significantly stronger large-scale magnetic fields than their partially convective (PC) counterparts with similar Rossby numbers ($R_{\textrm{o}}\approx1$), further suggesting differences in the dynamo mechanism between PC and FC stars \citep{See2025, Donati2023}. In this work, we present an analysis of spectropolarimetric observations of the planet hosting M dwarf star YZ~Ceti, and the discovery of an extreme change in its global magnetic field. 

\object{YZ~Ceti} (\object{GJ 54.1}) is a nearby M4.5\,V star located at a distance of $\sim3.7$ pc \citep{Gaia2023}. \cite{Schweitzer2019} report an effective temperature of $T_{\mathrm{eff}}=3151\pm51\,$K and a stellar mass of $M=0.142\pm0.010\,M_\odot$, placing YZ~Ceti near the low end of the M dwarf mass range and well below the theoretical $\sim0.35\,M_{\odot}$ transition to a fully convective interior \citep{Chabrier1997, Lu2024, Amard2019}. Multiple studies have inferred a $\sim67-69$\,d rotation period from ground-based time-series photometry \citep{Suarez2016, Engle2017, Jayasinghe2017, Diez2019}. \cite{Stock2020} measured a rotation period of $\sim 68.4$ days by combing V-band photometric observations from multiple telescopes.

YZ~Cet hosts three Earth-mass planets detected using radial velocity measurements \citep{Astudillo-Defru2017, Stock2020}. These planets: YZ~Cet~b, c, and d, have orbital periods $P_{\rm orb} = 2.02, 3.06, 4.66$ days and semi-major axes $r_{\rm orb} = 0.016, 0.02, 0.028$ au, respectively \citep{Stock2020}.  Strong radio emission was detected from this system only when the planet YZ~Cet~b was in two orbital sectors symmetric to the direction of Earth, suggesting a star-planet interaction scenario \citep{Trigilio2023, Pineda2023}. Assuming ECME, \cite{Trigilio2023} estimated the dipolar field strength of the M dwarf star to be $2.4$ kG. By comparing radiated and incident magnetic power in radio bands, the authors infer a planetary magnetosphere, estimating the lower limit of the magnetic field strength of YZ~Cet~b to be $\sim0.4$ G. 

This paper is structured as follows: we describe the observations performed in the near-infrared (IR) and optical bands in Section \ref{Observation}, report the detection of magnetic field and its variation in Section \ref{Analysis1}, magnetic imaging analysis in Section \ref{ZDI}, discuss our result and implications on planetary magnetic field calculation, and finally conclude in Section \ref{Discussion}.

\section{Observations} \label{Observation}

\subsection{CFHT: SPIRou} \label{SPIRou}

We acquired a total of 54 spectropolarimetric observations of YZ~Cet with the SpectroPolarimètre InfraRouge (SPIRou) mounted on the 3.6 m Canada–France–Hawaii Telescope (CFHT). SPIRou is a high-resolution near-infrared spectropolarimeter operating in the wavelength range 0.96 to 2 $\mu$m and has a resolving power of $R \sim 70,000$ \citep{Donati2020}. Each circular polarization observation with SPIRou consists of 4 sub-exposures associated with different orientations of the Fresnel rhomb retarders. The 4 sub-exposures are combined to produce one unpolarized (Stokes~$I$), one circularly polarized (Stokes~$V$) and two diagnostic null $N$ spectra \citep{Donati1997}. All SPIRou data were processed with the \textsc{APERO} SPIRou reduction pipeline (version 0.7.290, \citealt{Cook2022}).

\startlongtable
\begin{deluxetable*}{ccccccccc}
\tablecaption{Log of spectropolarimetric observations of YZ~Cet. Columns 1--9 list, respectively, the Modified Julian Date (MJD) at the mid point of the observation, the UTC date of observation, the rotation phase ($\phi_{\rm rot}$) with adopted ephemeris described in the text, the exposure time (in seconds), the peak signal-to-noise of the spectra, mean longitudinal field strength ($B_{\ell}$) in gauss { computed using integration limits of [-23,79] km/s}, corresponding error of the longitudinal field ($\sigma_{B_{\ell}}$), the false alarm probability (FAP), and detection flag of Least-Squares Deconvolution (LSD) Stokes V profile. The detection flag can be definite (DD), marginal (MD), or not detected (ND). \label{tab:obs_det}}
\tablehead{
\colhead{MJD} & \colhead{Date}& \colhead{$\phi_{\rm rot}$} & \colhead{Exposure (s)} & \colhead{SNR$_{\rm peak}$} & \colhead{$\langle B_{\ell} \rangle$ (G)} & \colhead{$\sigma_B$} (G) & \colhead{FAP} & \colhead{Det. Flag}
} 
\startdata 
\multicolumn{9}{c}{NARVAL Observation} \\
54705.119699 & 2008-09-18 & 0.68 & 3600 & 206 & -13 & 19 & 0.17 & ND \\
\multicolumn{9}{c}{ESPaDOnS Observations} \\
57652.338000 & 2016-09-21 & 0.00 & 579 & 217 & 129 & 28 & 0.0 & DD \\
57673.457000 & 2016-10-12 & 0.30 & 537 & 217 & 48 & 27 & 0.29 & ND \\
58004.514000 & 2017-09-08 & 0.05 & 1120 & 305 & -79 & 19 & 2.6e-13 & DD \\
58124.297000 & 2018-01-06 & 0.77 & 1104 & 276 & 294 & 30 & 0.0 & DD \\
\multicolumn{9}{c}{SPIRou Observations (epoch 2023)} \\
60244.330366 &  2023-10-27 &  0.20 &  159 &  344 &  128 &  17 &  6.9e-14 &  DD \\
60245.357390 &  2023-10-28 &  0.21 &  165 &  337 &  110 &  19 &  4.7e-11 &  DD \\
60246.410315 &  2023-10-29 &  0.23 &  166 &  330 &  153 &  23 &  5.6e-10 &  DD \\
60247.482963 &  2023-10-30 &  0.24 &  168 &  330 &  56 &  14 &  2.4e-07 &  DD \\
60248.372378 &  2023-10-31 &  0.25 &  166 &  310 &  68 &  15 &  2.1e-07 &  DD \\
60249.394604 &  2023-11-01 &  0.27 &  166 &  329 &  48 &  16 &  7.5e-12 &  DD \\
60250.388063 &  2023-11-02 &  0.28 &  164 &  341 &  90 &  19 &  5.3e-15 &  DD \\
60251.415297 &  2023-11-03 &  0.30 &  167 &  347 &  76 &  18 &  1.1e-15 &  DD \\
60252.323407 &  2023-11-04 &  0.31 &  167 &  349 &  73 &  19 &  3.0e-14 &  DD \\
60253.302499 &  2023-11-05 &  0.32 &  163 &  330 &  17 &  18 &  3.0e-12 &  DD \\
60254.386517 &  2023-11-06 &  0.34 &  163 &  270 &  22 &  21 &  3.2e-08 &  DD \\
60255.306807 &  2023-11-07 &  0.35 &  165 &  209 &  1 &  28 &  1.1e-01 &  ND \\
60256.402831 &  2023-11-08 &  0.37 &  163 &  318 &  54 &  17 &  3.9e-08 &  DD \\
60265.272345 &  2023-11-17 &  0.50 &  165 &  354 &  39 &  14 &  4.0e-03 &  ND \\
60266.320913 &  2023-11-18 &  0.51 &  160 &  310 &  32 &  14 &  2.0e-02 &  ND \\
60268.323592 &  2023-11-20 &  0.54 &  165 &  208 &  57 &  36 &  3.0e-02 &  ND \\
60268.337532 &  2023-11-20 &  0.54 &  164 &  237 &  71 &  21 &  8.6e-04 &  MD \\
60269.299946 &  2023-11-21 &  0.55 &  165 &  283 &  53 &  16 &  1.0e-05 &  MD \\
60270.250431 &  2023-11-22 &  0.57 &  166 &  311 &  64 &  12 &  5.3e-09 &  DD \\
60271.237809 &  2023-11-23 &  0.58 &  168 &  312 &  58 &  13 &  1.1e-07 &  DD \\
60272.210162 &  2023-11-24 &  0.60 &  169 &  348 &  58 &  11 &  4.5e-14 &  DD \\
60273.231378 &  2023-11-25 &  0.61 &  164 &  349 &  85 &  13 &  0.0 &  DD \\
60274.195094 &  2023-11-26 &  0.62 &  164 &  333 &  80 &  12 &  0.0 &  DD \\
60283.224272 &  2023-12-05 &  0.75 &  163 &  346 &  70 &  11 &  0.0 &  DD \\
60284.206427 &  2023-12-06 &  0.77 &  165 &  325 &  86 &  12 &  0.0 &  DD \\
60285.184311 &  2023-12-07 &  0.78 &  168 &  342 &  87 &  13 &  5.6e-16 &  DD \\
60286.185006 &  2023-12-08 &  0.80 &  165 &  259 &  90 &  20 &  1.0e-07 &  DD \\
60288.313951 &  2023-12-10 &  0.83 &  165 &  320 &  64 &  13 &  2.7e-05 &  MD \\
60289.292561 &  2023-12-11 &  0.84 &  165 &  344 &  95 &  14 &  1.7e-10 &  DD \\
60290.246338 &  2023-12-12 &  0.85 &  166 &  338 &  96 &  14 &  3.3e-13 &  DD \\
60291.309143 &  2023-12-13 &  0.87 &  165 &  305 &  117 &  16 &  1.1e-16 &  DD \\
60292.307473 &  2023-12-14 &  0.88 &  168 &  329 &  118 &  15 &  0.0 &  DD \\
60293.246040 &  2023-12-15 &  0.90 &  167 &  330 &  169 &  16 &  0.0 &  DD \\
60294.321964 &  2023-12-16 &  0.91 &  161 &  342 &  174 &  16 &  0.0 &  DD \\
60295.233091 &  2023-12-17 &  0.93 &  163 &  340 &  214 &  18 &  0.0 &  DD \\
60302.201034 &  2023-12-24 &  0.03 &  166 &  316 &  199 &  21 &  0.0 &  DD \\
60303.201699 &  2023-12-25 &  0.04 &  168 &  323 &  200 &  21 &  0.0 &  DD \\
60304.190691 &  2023-12-26 &  0.05 &  159 &  348 &  177 &  19 &  0.0 &  DD \\
60305.190698 &  2023-12-27 &  0.07 &  163 &  320 &  205 &  23 &  0.0 &  DD \\
60306.191436 &  2023-12-28 &  0.08 &  163 &  344 &  170 &  21 &  0.0 &  DD \\
60307.191251 &  2023-12-29 &  0.10 &  167 &  319 &  140 &  23 &  1.1e-07 &  DD \\
60333.198239 &  2024-01-24 &  0.47 &  166 &  339 &  34 &  22 &  7.6e-01 &  ND \\
60334.216074 &  2024-01-25 &  0.49 &  166 &  255 &  83 &  33 &  2.5e-01 &  ND \\
\multicolumn{9}{c}{SPIRou Observations (epoch 2024)} \\
60571.485287 &  2024-09-18 &  0.89 &  218 &  369 &  302 &  24 &  0.0 &  DD \\
60577.552375 &  2024-09-24 &  0.98 &  216 &  379 &  416 &  24 &  0.0 &  DD \\
60594.432235 &  2024-10-11 &  0.22 &  213 &  288 &  152 &  26 &  7.3e-11 &  DD \\
60595.372312 &  2024-10-12 &  0.23 &  218 &  389 &  108 &  18 &  0.0 &  DD \\
60602.408640 &  2024-10-19 &  0.33 &  214 &  368 &  -24 &  14 &  3.6e-01 &  ND \\
60608.345174 &  2024-10-25 &  0.42 &  218 &  347 &  -96 &  15 &  0.0 &  DD \\
60621.461839 &  2024-11-07 &  0.61 &  213 &  341 &  -100 &  16 &  0.0 &  DD \\
60627.409250 &  2024-11-13 &  0.69 &  214 &  340 &  -41 &  16 &  1.3e-06 &  DD \\
60635.212536 &  2024-11-21 &  0.81 &  213 &  376 &  120 &  20 &  2.6e-06 &  DD \\
60640.277446 &  2024-11-26 &  0.88 &  215 &  373 &  262 &  21 &  0.0 &  DD \\
60654.258253 &  2024-12-10 &  0.08 &  217 &  322 &  325 &  23 &  0.0 &  DD 
\enddata
\end{deluxetable*}

The SPIRou observations were acquired over two observing epochs in 2023 and 2024. Both observing epochs covered the $\sim68$\,d rotational period inferred from ground-based photometry \citep{Stock2020}. The first 43 observations were carried out between 27 October 2023 and 25 January 2024 (Proposal ID: 23BD01, PI: J. Morin) spanning a period of 89\,d with a median temporal separation of 1\,d. The median peak signal-to-noise ratio (SNR) of the 2023 spectra is $\sim330$ per spectral pixel.

\begin{figure*}
     \centering
     \begin{subfigure}[b]{0.32\textwidth}
         \centering
         \includegraphics[width=\textwidth]{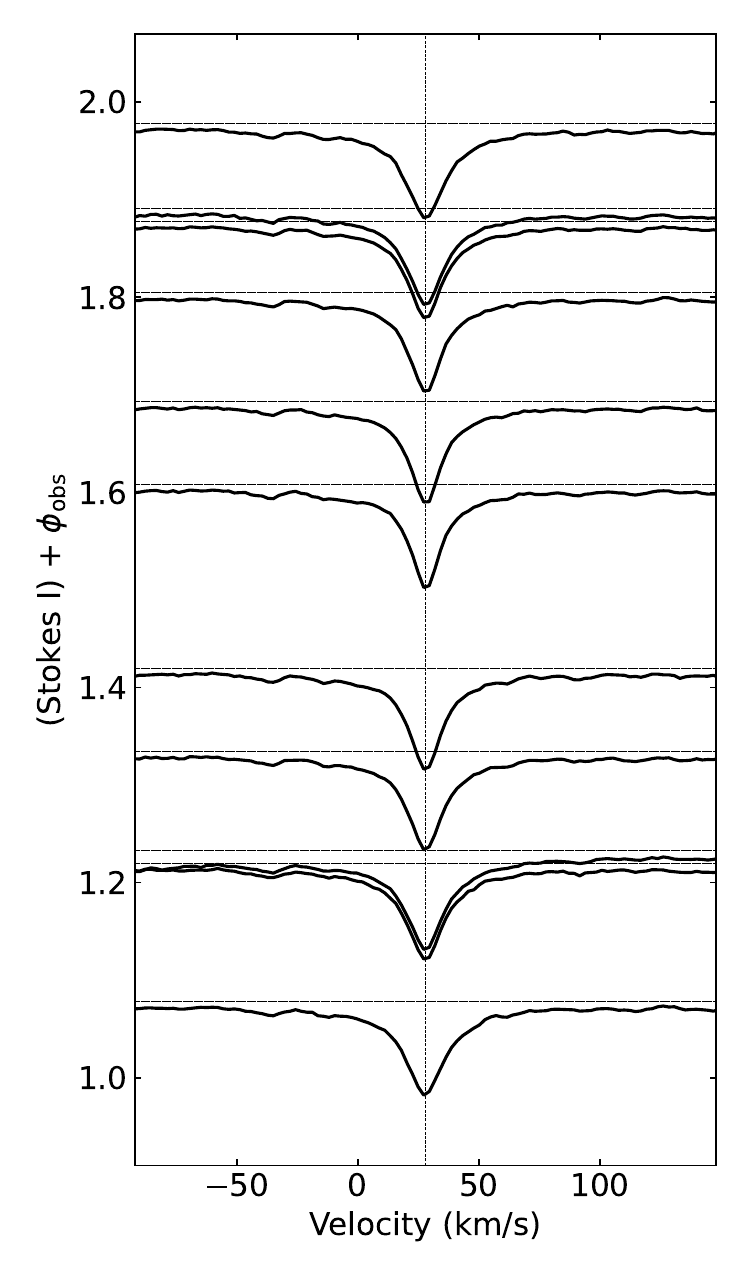}
         \caption{Stokes I}
         \label{fig:StokesI}
     \end{subfigure}
     \begin{subfigure}[b]{0.32\textwidth}
         \centering
         \includegraphics[width=\textwidth]{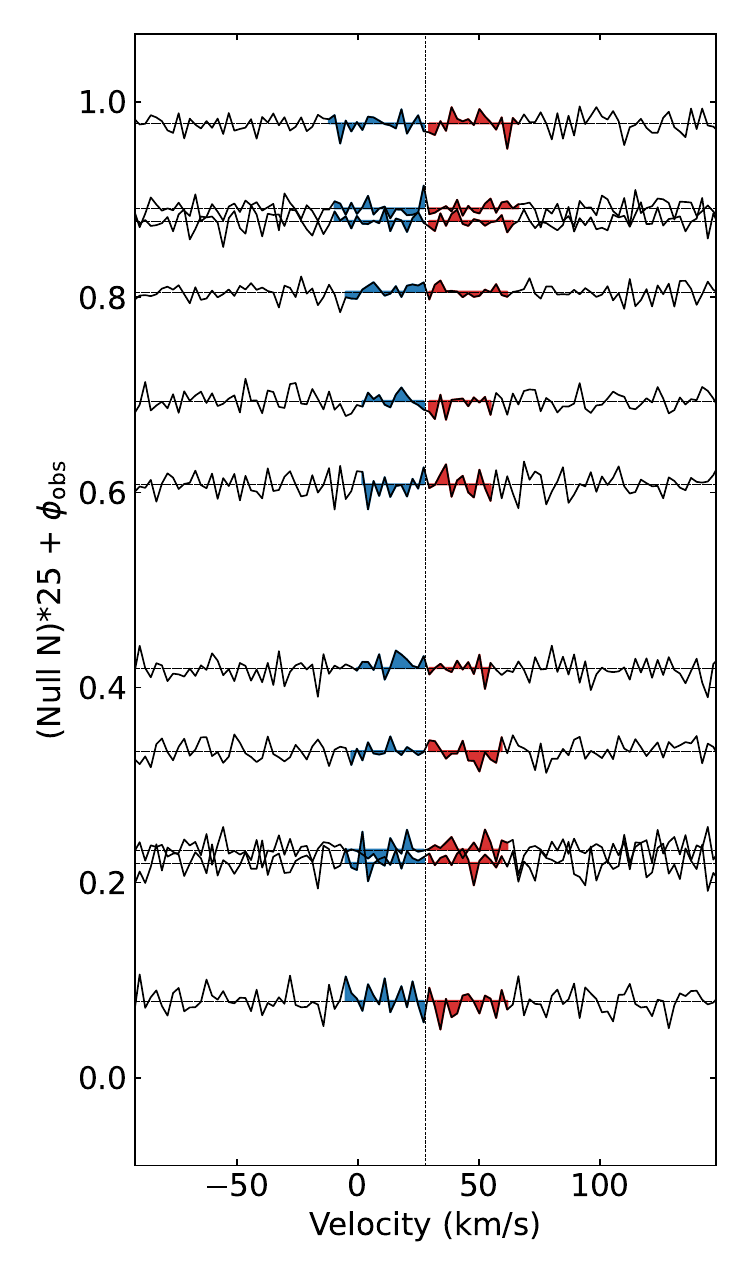}
         \caption{Null N}
         \label{fig:StokesN}
     \end{subfigure}
     \begin{subfigure}[b]{0.32\textwidth}
         \centering
         \includegraphics[width=\textwidth]{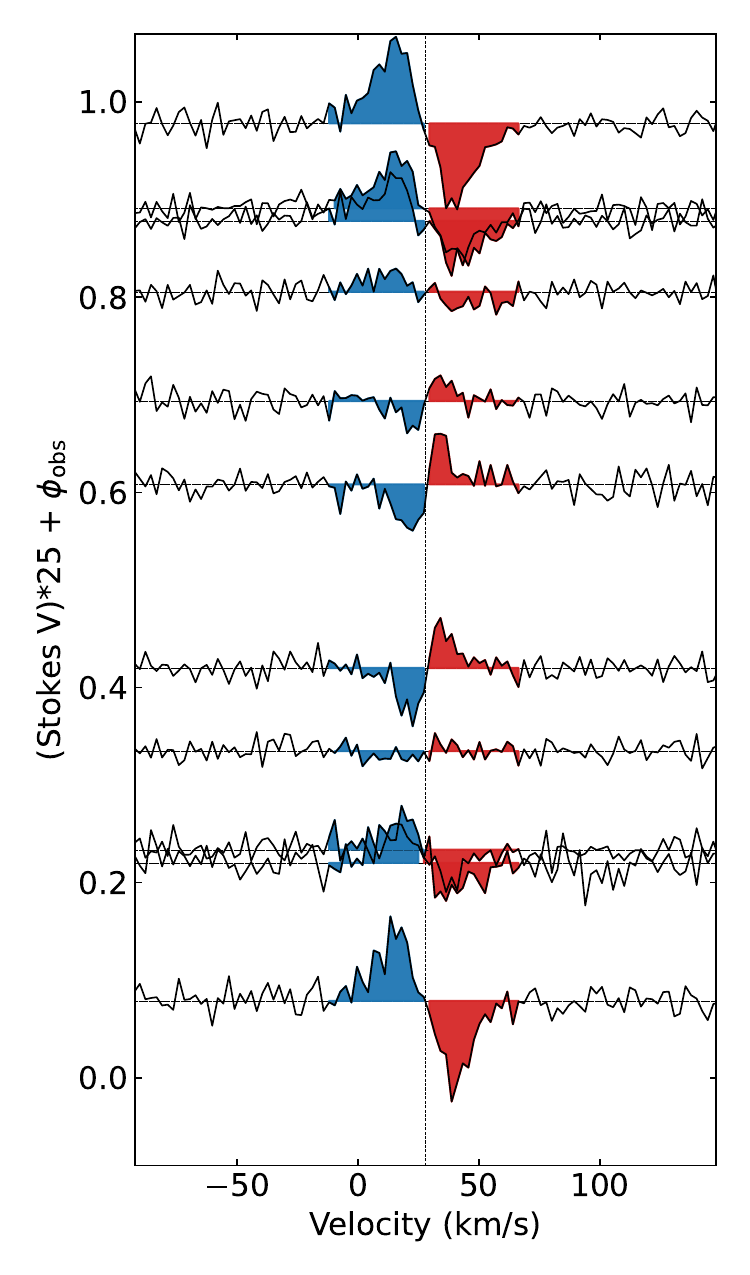}
         \caption{Stokes V}
         \label{fig:StokesV}
     \end{subfigure}
        \caption{Timeseries of LSD profiles of YZ~Cet derived from the 2024 SPIRou observations. The three panels show (a) Stokes I, (b) Null N, and (c) Stokes V, respectively. The profiles are shifted vertically by an offset corresponding to their rotation phases according to the ephemeris given in the text. The vertical dashed line represent the mean radial velocity obtained from the LSD I profiles. For panel (b) and (c), the shaded regions represent the integration limits while determining the longitudinal fields. \label{fig:Stokes_AB}}
\end{figure*}

The next 11 SPIRou observations were obtained at near uniform cadence between between 18 September 2024 and 10 December 2024 (Proposal ID: 24BC28, PI: A. Biswas). The 2024 observations span $83$\,d with a median temporal separation of $7$\,d. The median peak SNR of the 2024 spectra is $\sim370$ per spectral pixel. The log of our observations is presented in Table \ref{tab:obs_det}.

\subsection{Archival ESPaDOnS and NARVAL Observations} \label{archival}

We identified 4 archival optical spectropolarimetric observations obtained with the ESPaDOnS (Echelle Spectropolarimetric Device for the Observation of Stars) spectropolarimeter located at CFHT \citep{Donati2006_espadons}. ESPaDOnS operates in the wavelength range of 370 to 1050 nm at a resolution of R = 65000. These observations were taken under the `SPIRou input catalog' project \citep{Moutou2017}. The first two observations were taken in 2016 under proposal ID 16B27 (PI: L. Malo), and the other two observations were taken in 2017--18 under proposal ID 17BD98 (PI: C. Moutou). 

From the PolarBase database \citep{Petit2014} we retrieved one archival observation of YZ~Ceti obtained with the NARVAL spectropolarimeter in 2008. NARVAL was a twin instrument of ESPaDOnS mounted on Télescope Bernard Lyot (TBL) at Pic du Midi Observatory. A log of all archival observations is presented in Table \ref{tab:obs_det}.

\begin{figure*}
     \centering
     \begin{subfigure}[b]{0.62\textwidth}
         \centering
         \includegraphics[width=\textwidth]{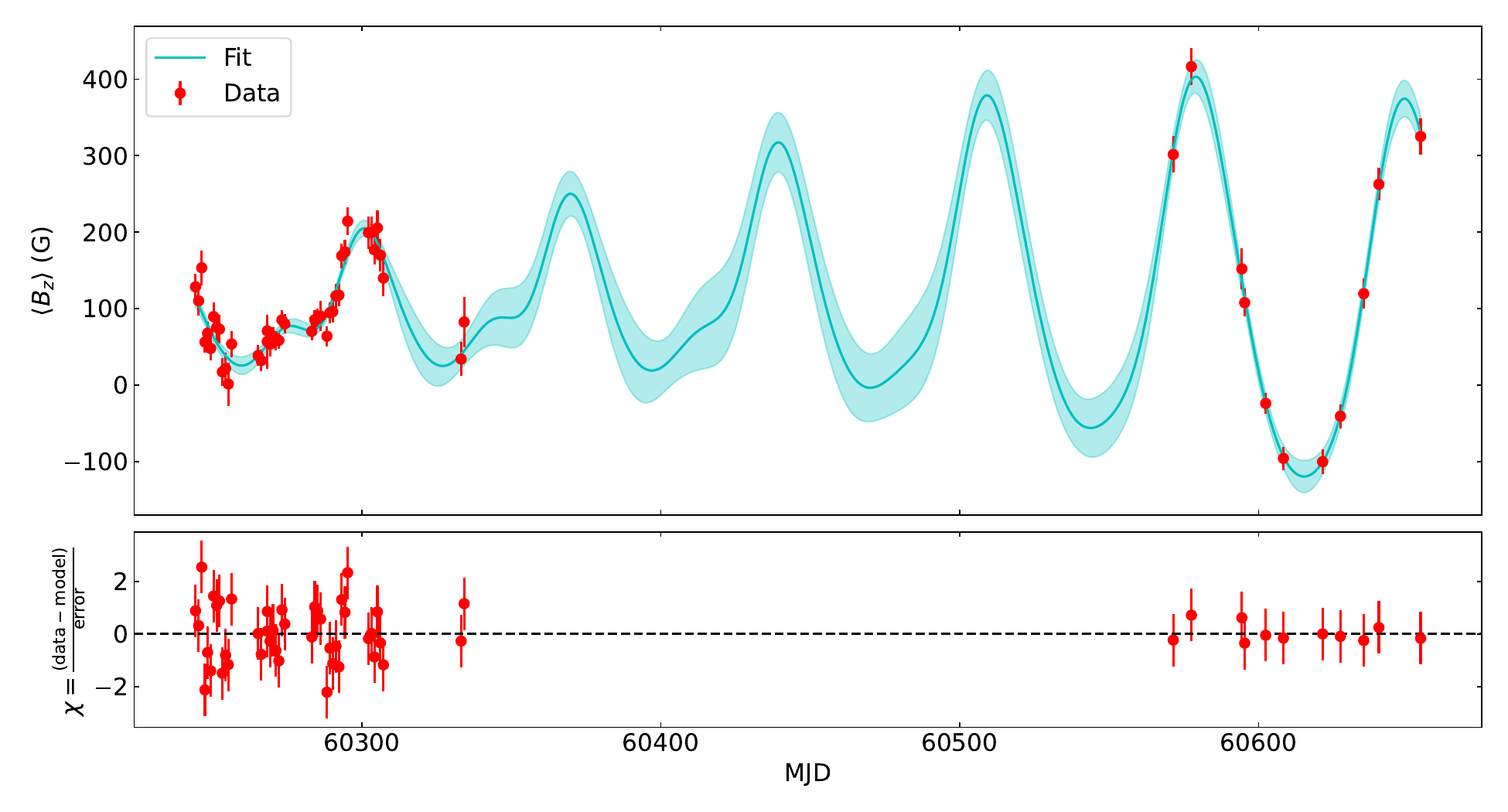}
         \caption{Temporal variation and GRP fit}
         \label{fig:GLP}
     \end{subfigure}
     \begin{subfigure}[b]{0.36\textwidth}
         \centering
         \includegraphics[width=\textwidth]{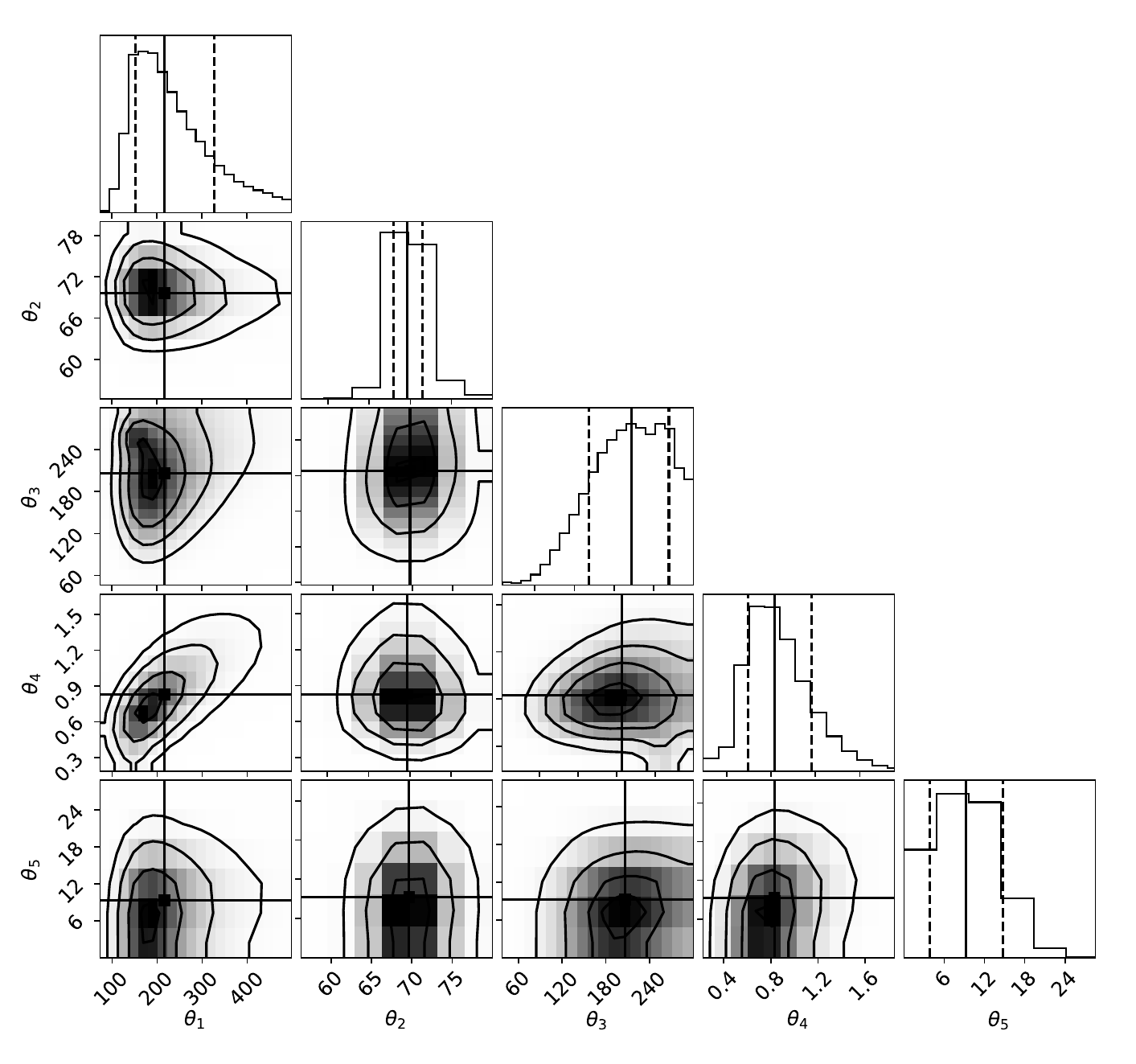}
         \caption{MCMC posterior distribution}
         \label{fig:GRP_mcmc}
     \end{subfigure}
        \caption{GRP fit: (a) Gaussian regression process fit to the SPIRou data. The red data in the top column correspond to the longitudinal field strength in gauss. The blue line represent the best fit, with the shaded region as the fit uncertainty. The bottom column represent the residual, defined as $\chi$ = residual/error. The fit has a reduced $\chi^2$ of $\approx1.1$. All data points are within $2\sigma$ limit of the GRP fit. (b) The posterior density distribution resulting from MCMC analysis of GRP model of $B_{\ell}$. The contours represent the $1\sigma$, $2\sigma$, and $3\sigma$ levels. The black solid lines mark the median values of the posterior distribution function (PDF). The black dashed lines represent the 16 per cent and 84 per cent percentiles of the PDF.  \label{fig:GRP}}
\end{figure*}

\section{Magnetic Measurements and Variability}  \label{Analysis1}

\subsection{LSD Profiles} \label{LSD}

We applied the Least-Squares Deconvolution (LSD; \citealt{Donati1997}) technique to each co-added spectrum. The LSD procedure combines a subset of photospheric atomic lines and gives a high SNR mean unpolarized profile (LSD Stokes I), a mean circular polarization profile (LSD Stokes V), and a mean diagnostic null profile (LSD N). We used the Python-based open-source implementation of the LSD algorithm \texttt{LSDpy}\footnote{\url{https://github.com/folsomcp/LSDpy}} within the \texttt{SpecpolFlow}\footnote{\url{https://folsomcp.github.io/specpolFlow/}} software package \citep{Folsom2025}. 

To create the LSD line mask, we extracted a line list from the Vienna Atomic Line Database ({\textsc{VALD3}}, \citealt{Piskunov1995, Kupka1999}), for the stellar parameters $T_{\rm eff} = 3500$ K, $\log g = 5.0$, { solar metallicity, and a VALD estimated depth of at least 0.1. Only atomic lines were used, and molecular lines were rejected.  We then manually filtered the VALD line list, only selecting lines  where the LSD model spectrum  well reproduced the observed Stokes~I spectrum. More specifically, we compared the model spectrum generated by \texttt{LSDpy} with the observed spectrum using the interactive \texttt{cleanMaskUI} tool from the \texttt{SpecpolFlow} package, and iteratively recomputed LSD models after removing a significant number of poorly fit lines. Lines in regions with strong telluric contamination were excluded, since the telluric correction becomes unreliable in these cases. Lines with significant molecular blending were excluded.  Specifically, lines in the observation where there appeared to be a blending feature that was not reproduced in the LSD model, where that feature was greater than 20\% of the line depth, were considered likely to be blended with significant molecular features and excluded. Finally, lines with large Lorentzian wings were also excluded (7 lines with depth estimates above 0.5), likely produced by van der Waals broadening in strongly saturated lines, since they do not satisfy the assumption in LSD that all lines have similar shapes. This left us with a mask that contained only 45 high-quality lines between 1040 and 1310 nm.} While this restrictive approach to generating the line mask somewhat reduced the effective SNR of our LSD profiles, it minimized the systematic distortions to the LSD profiles, allowing us to produce the highest quality magnetic maps possible. 

Depths for the lines in the mask were taken from VALD's depth estimate, but modified where necessary following a procedure similar to \citet{Grunhut2017}. We used \texttt{cleanMaskUI} tool from the \texttt{SpecpolFlow} to identify lines where the LSD model disagreed with observed line depths by more than 20\%.  The depths for these lines were then fit using \texttt{cleanMaskUI}, assuming that the depths of the remaining lines were correct.  This assumption is necessary to split the degeneracy between line depths and the Stokes I LSD profile amplitude.  This resulted in depth corrections to 27 out of the 45 lines.

For SPIRou spectra, the LSD profiles were calculated adopting a normalization wavelength $\lambda_0 = 1200$ nm and an effective Land\'e factor $g_{\rm eff} = 1.2$, which are close to the weighted average values of the corresponding parameters of the lines included in our custom mask. For ESPaDOnS and NARVAL, we use $\lambda_0=550$ nm with the same $g_{\rm eff}$. The time series of LSD Stokes $I$, $V$, and null $N$ profiles computed from the 2024 SPIRou observations are shown in Fig. \ref{fig:Stokes_AB}.

We calculated false alarm probabilities (FAP, \citealt{Donati1992}) based on $\chi^2$ statistics to diagnose how robustly a signal is detected over the noise level in the computed LSD profiles. We define a detection to be definite if the FAP value is below $10^{-5}$. We define a marginal detection for which $10^{-5}<$FAP$<10^{-3}$. Finally, for FAP$>10^{-3}$, we consider it a non-detection. From the 2023 SPIRou Stokes V observations, we obtained 7 non-detections, 5 marginal detections, and 31 definite detections. From the SPIRou Stokes V observations in 2024, we obtain 10 definite detections and one non-detection. All 4 ESPaDOnS spectra show clear Stokes V detections, while the archival NARVAL spectra resulted in a non-detection.

\begin{figure*}
     \centering
     \begin{subfigure}[b]{0.9\textwidth}
         \centering
         \includegraphics[width=\textwidth]{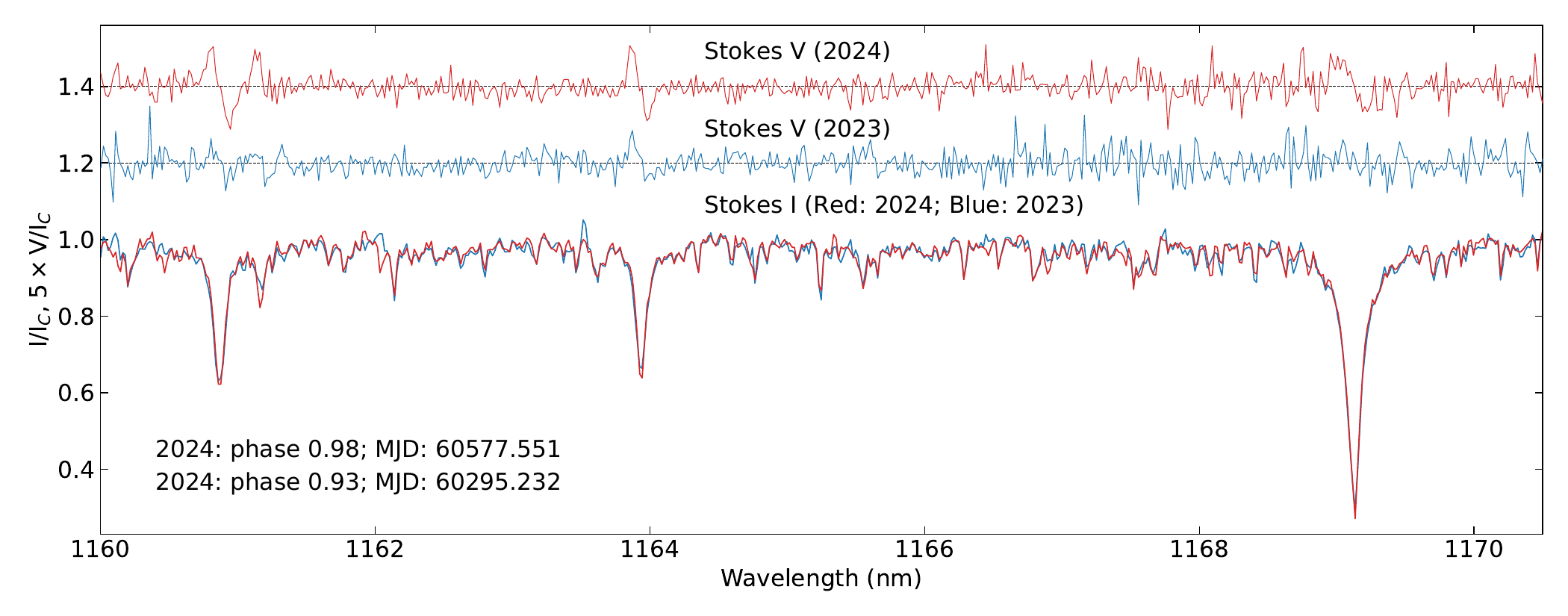}
         \caption{Comparison of 2023 and 2024 SPIRou spectra around $\phi_{\rm rot} \sim0$.}
         \label{fig:4a}
     \end{subfigure}
     \begin{subfigure}[b]{0.42\textwidth}
         \centering
         \includegraphics[width=\textwidth]{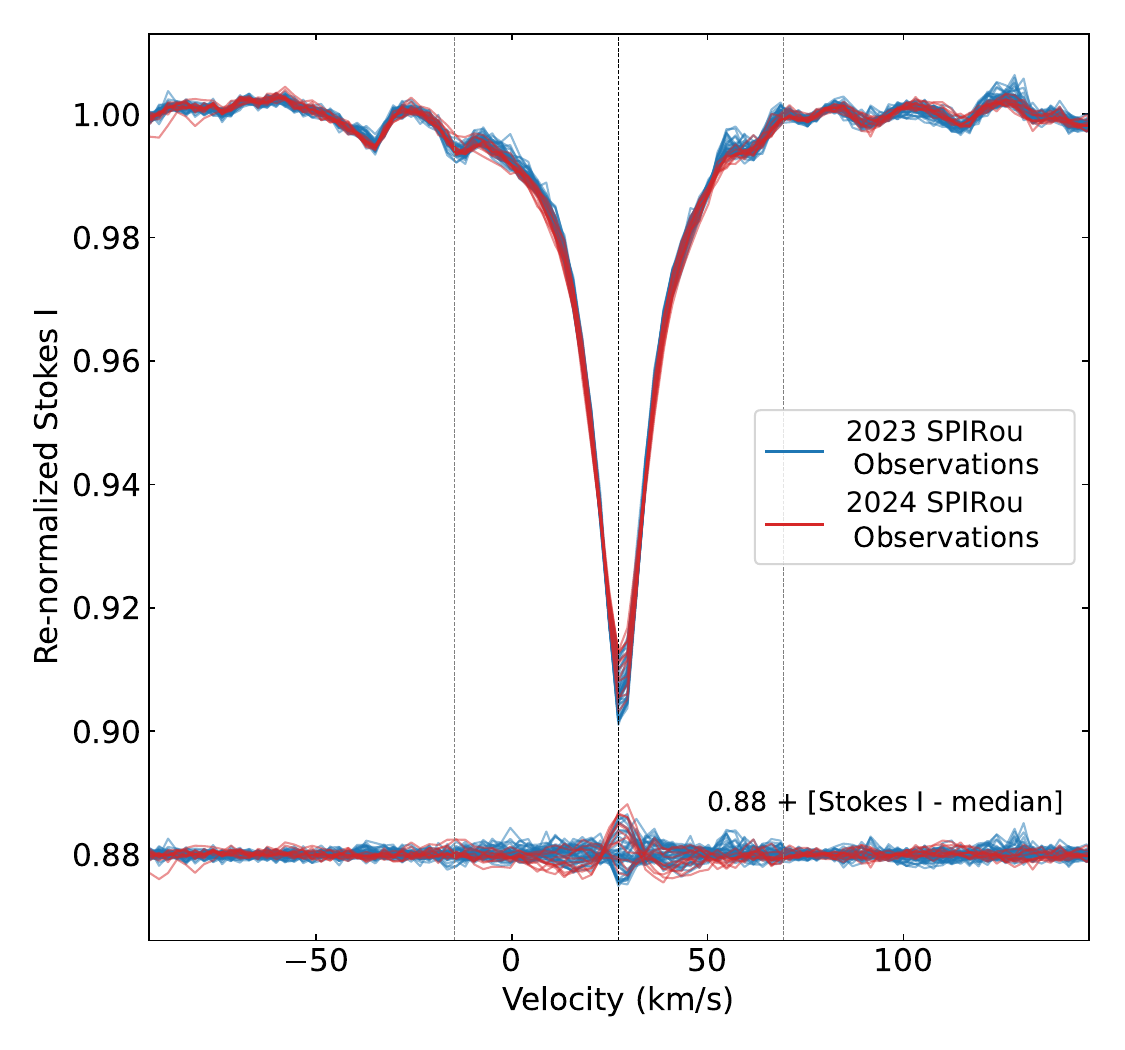}
         \caption{Comparison of LSD Stokes I}
         \label{fig:4b}
     \end{subfigure}
     \begin{subfigure}[b]{0.42\textwidth}
         \centering
         \includegraphics[width=\textwidth]{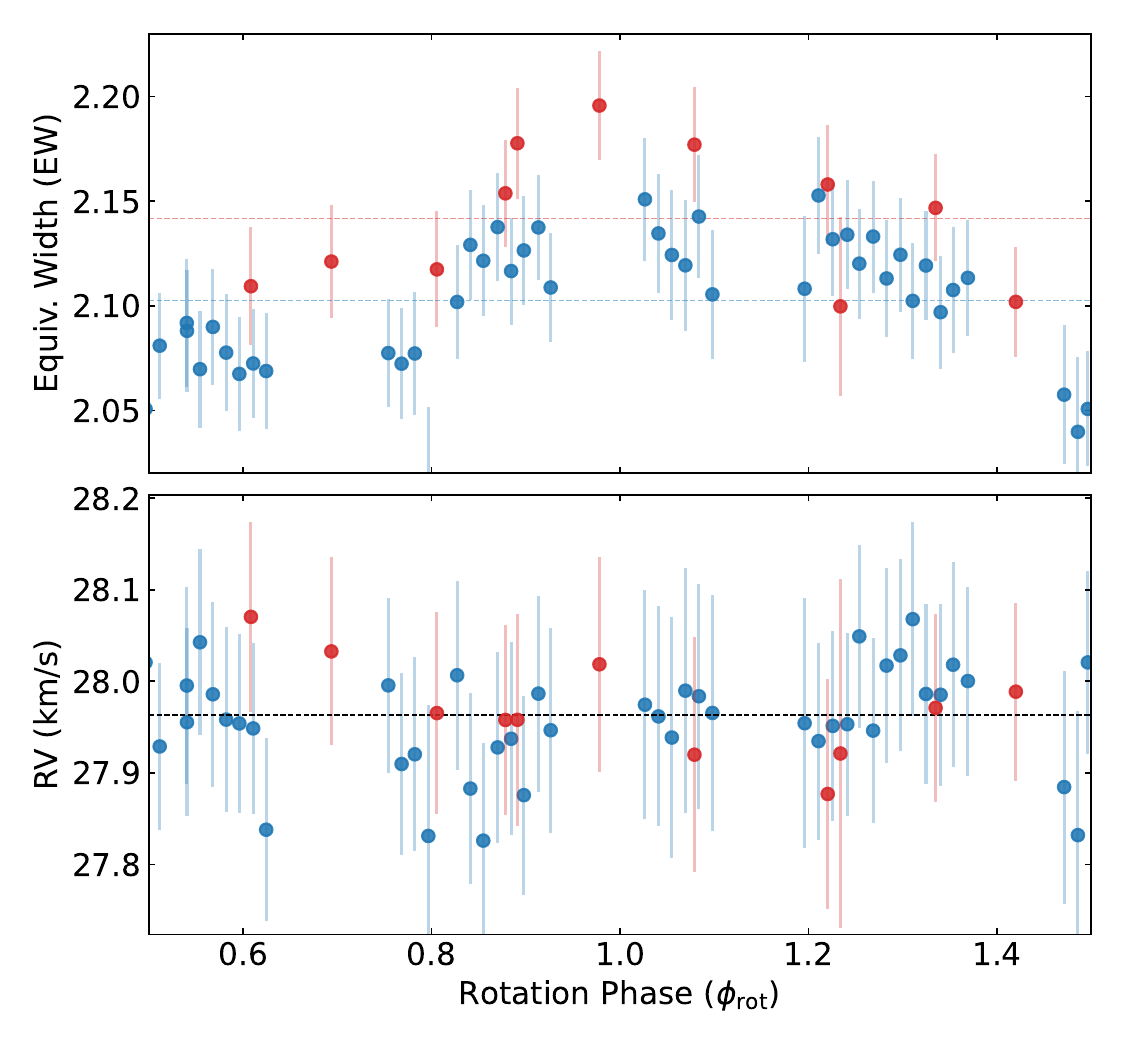}
         \caption{Comparison of EW and RV}
         \label{fig:4c}
     \end{subfigure}
        \caption{Comparison of SPIRou spectra between 2023 (blue) and 2024 (red) observations. (a) A small wavelength interval of Stokes I and Stokes V (shifted) spectra near rotation phase $\phi_{\rm rot} \sim0$ showing the change in Stokes V amplitude over time. Both spectra have comparable SNR. (b) Comparison of LSD Stokes I spectra of 2023 and 2024 observations. The difference between individual observed LSD Stokes I profiles and the median observed profile is shown (shifter by 0.88) to illustrate the lack of significant rotational variability, and to justify the assumption of uniform brightness for further ZDI analysis. The vertical dashed line represents the velocity range for equivalent width (EW) calculation. (c) Rotational variation of EW and radial velocity (RV) in the 2023 and 2024 observations. The mean EW for the 2023 and 2024 observations are shown as blue and red dashed horizontal dashed lines, respectively. In the bottom panel, the mean RV is shown with black dashed line.  \label{fig:SPIRou_spectra}}
\end{figure*}

We find some clear differences between Stokes V LSD profiles from the 2023 and 2024 observations.  The 2023 Stokes V profiles have a lower maximum amplitude than the 2024 profiles.  The 2023 profiles also show more complexity at some phases, while the 2024 profiles show a stronger reversal of sign.  Fig \ref{fig:4a} illustrates this, comparing Stokes V LSD profiles from similar rotation phases (with the ephemeris derived in Sect.\ \ref{Bz_vary}).  These differences is Stokes V profile morphology provide an early hint that that the magnetic field may have changed substantially between 2023 and 2024.

\subsection[Bl Variation]{$B_{\ell}$ Variation} \label{Bz_vary}

The mean longitudinal magnetic field $B_{\ell}$ was computed for each observation from the LSD Stokes $I$ and $V$ profiles according to the formula \citep{Donati1997, Wade2000}:

\begin{equation}
 B_{\ell} = -\frac{2.14 \times 10^{11}}{\lambda_0 g_{\text{eff}} \ c}\frac{\int v V(v) dv}{\int \left[I_c - I(v)\right] dv} ,
\end{equation}

\noindent
where $B_{\ell}$ is in gauss (G), $v$ is the radial velocity in km/s, $\lambda_0$, in nm is the mean wavelength in the LSD profile, $c$ is the speed of light in vacuum in km/s, $g_{\rm eff}$ is the mean effective Land\'e factor used for normalization of the LSD weights, and $I_c$ is the continuum level (1, in case of normalized spectra). The measured values of $B_{\ell}$, and the associated uncertainties for the SPIRou observations are listed in Table \ref{tab:obs_det}. The temporal variation of $B_{\ell}$ is shown in Fig. \ref{fig:GLP}. 

{ As $B_{\ell}$ may evolve over a few rotational cycles,} we used Gaussian process regression (GPR) analysis of the $B_{\ell}$ data instead of a standard Fourier analysis, following the procedure described by \cite{Donati2023} or \cite{Haywood2014}. We use a quasi-periodic (QP) kernel whose covariance function $c(t,t')$ is defined as:

\begin{eqnarray}
c(t,t') = \theta_1^2 \exp \left( -\frac{(t-t')^2}{2 \theta_3^2} -\frac{\sin^2 \left( \frac{\pi (t-t')}{\theta_2} \right)}{2 \theta_4^2} \right) 
\label{eq:covar}
\end{eqnarray}

\noindent
where $\theta_1$ is the GP amplitude (in G), $\theta_2$ the rotation period ($P_{\rm rot}$, in days), $\theta_3$ the evolution timescale of $B_{\ell}$ (in days), and $\theta_4$ is a smoothing parameter. The log-likelihood $\mathcal{L}$ is computed as:

\begin{multline}
      2 \log \mathcal{L} = -n \log(2\pi) - \log|C+\Sigma+S| \\ - y^T (C+\Sigma+S)^{-1} y,  \label{eq:llik}
\end{multline}

\begin{figure}
    \centering
    \includegraphics[width=0.99\linewidth]{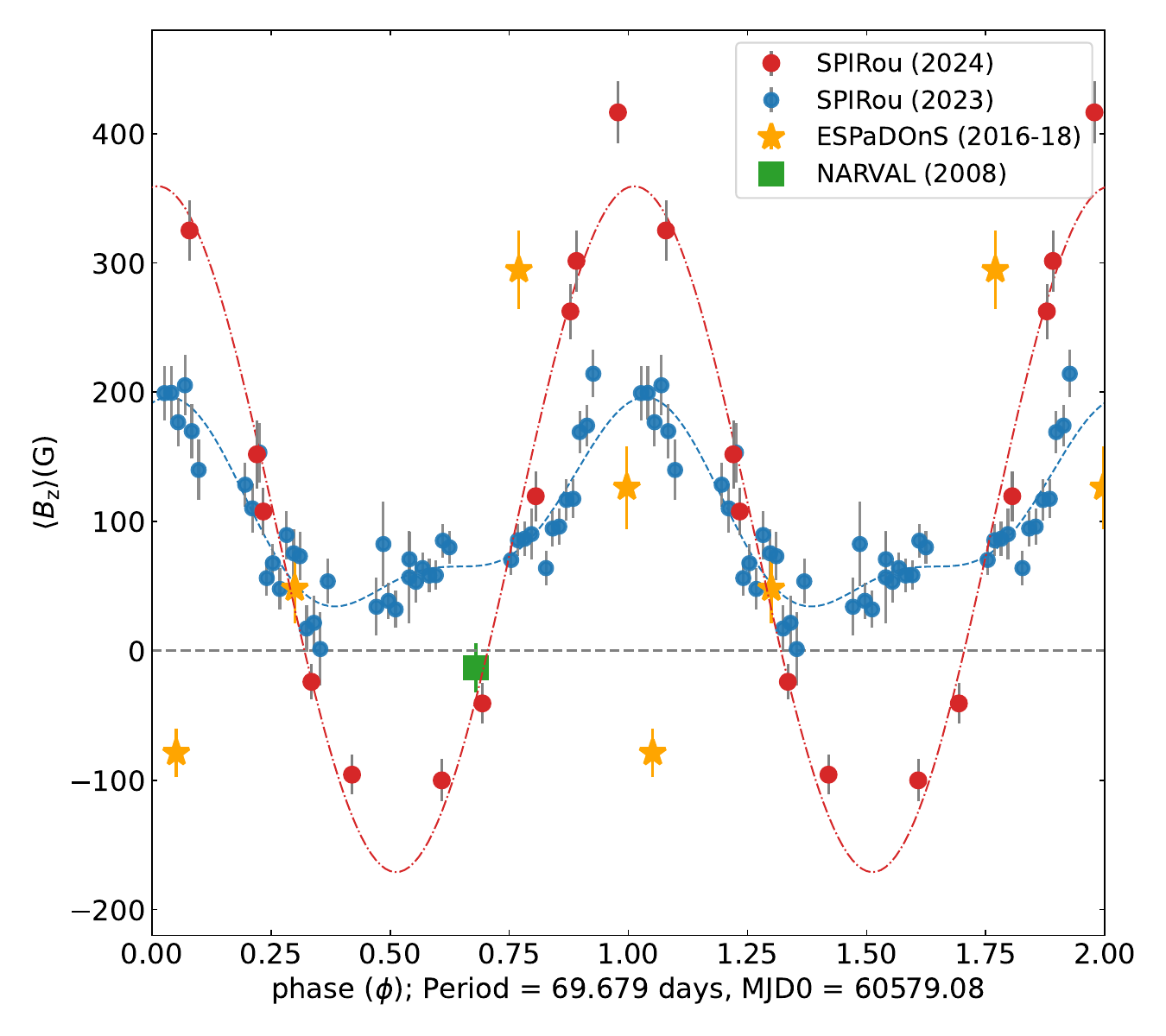}
    \caption{Phase folded $B_{\ell}$ variation of the 2023 SPIRou observations (in blue), 2024 SPIRou observations (in red), ESPaDOnS observations (in orange), and NARVAL observation (in green). For the 2024 observations, we fit a single sinusoid with the rotational frequency, while for the 2023 SPIRou observations we fit an additional first harmonic. All data were folded with the same ephemeris, as described in the text.}
    \label{fig:all_Bz}
\end{figure}

\noindent
where $C$ is the covariance matrix, $\Sigma$ the diagonal variance matrix of $B_{\ell}$, and $S = \theta_5^2 I$ represents the added white noise. To estimate parameter correlations and uncertainties, we perform a Markov Chain Monte Carlo (MCMC) analysis \citep{Foreman-Mackey2013}. We restrict the value of $\theta_3$ to be greater than $\theta_2$; start with a uniform prior for $\theta_2$ in the range [65,80] as suggested from the Lomb-Scargle periodogram of the $B_{\ell}$ data. For the smoothing parameter, we set a uniform prior in the range [0,3], and for both $\theta_1$ and $\theta_5$, we use modified Jeffreys priors. We determine the rotation period as the value having the largest likelihood. For the main analysis, we only used SPIRou data, as the archival ESPaDOnS observations are separated by a large time gap. To validate rotation periods, following \cite{Donati2023}, we compare the GPR model with a baseline where $\theta_2=1500$~d, $\theta_3=300$~d, and $\theta_4=1$, assessing the difference in the marginal log-likelihood $\Delta \log \mathcal{L}_M$. We classify results as definite detections ($\Delta \log \mathcal{L}_M>10$), marginal detections ($5<\Delta \log \mathcal{L}_M \leq 10$), or non-detections. For our data, we find $\Delta \log \mathcal{L}_M \approx 56$, which indicates a robust detection of rotational modulation. The final reduced $\chi^2$ was $\approx1.1$. The GRP fit to the SPIRou data is shown in Fig.\ref{fig:GLP}. The results of the MCMC analysis are shown in Fig. \ref{fig:GRP_mcmc}. We also tried to include the ESPaDOnS observations in the GRP fit, which gives a similar value of the rotation period with higher uncertainty. So, for all further analysis, we include the results obtained from the SPIRou measurements only.

The median best-fit values with 68\% confidence obtained from the MCMC analysis of the SPIRou data are: amplitude $\theta_1 = 220_{-64}^{+107}$ G, rotation period $\theta_2 = 69.7_{-1.7}^{+1.8}$ days, evolution timescale $\theta_3 = 204_{-63}^{+58}$ days, $\theta_4 = 0.85_{-0.24}^{+0.36}$, and reference MJD, MJD$_0 = 60579.08 \pm 0.33$. The amplitude has large uncertainties because of an extreme change of the overall amplitude between the two observing epochs. The evolution timescale is also shorter than 3 rotation cycles, suggesting a rapid evolution of the magnetic field. In the following analysis (and also in Table \ref{tab:obs_det}), all data are phased according to the following ephemeris:

\begin{equation}
    {\rm MJD} = {\rm MJD}_0 + P_{\rm rot} E,
\end{equation}

\noindent
where MJD$_0 = 60579.08$, and $P_{\rm rot} = 69.7$ days. This $P_{\rm rot}$ is used as input for our ZDI analysis in Sec. \ref{ZDI}. 

A phase-folded $B_{\ell}$ curve for all magnetic measurements is shown in Fig. \ref{fig:all_Bz}. For the 2024 SPIRou data, the $B_{\ell}$ variation appears approximately sinusoidal, consistent with a predominantly dipolar field geometry. In contrast, the 2023 data deviate from a sinusoidal trend, suggesting a more complex magnetic morphology. The amplitude of the 2024 $B_{\ell}$ curve is much larger than the 2023 curve, suggesting the dipolar component of the magnetic field has strengthened substantially in that time.  Additionally, the 2024 $B_{\ell}$ curve crosses zero to show negative values, while the 2023 curve remains positive at all phases.  This suggests a change in the obliquity of the dipole component (the angle between the dipole and rotation axis), although a large decrease in the quadrupole component strength could also influence this $B_{\ell}$ behavior.  These large changes between datasets separated by $\sim$300 days are consistent with the GRP evolution time scale ($\theta_3$) of $204_{-63}^{+58}$ days.

\subsection{Extreme change of the Stokes V spectrum}

We have already shown a possible major change in the global magnetic field from the $B_{\ell}$ measurements in Fig. \ref{fig:all_Bz}. In Fig. \ref{fig:SPIRou_spectra}, we investigate the change in individual spectra between the 2023 and 2024 observations. We reiterate that all data were analyzed with the same normalization and LSD extraction procedures. To compare a similar rotation phase with the highest SNR in Stokes V, we compare two spectra near rotation phase $\phi_{\rm rot} \sim0$ (Fig. \ref{fig:4a}). Although the Stokes~V signatures were clearly detected in individual lines of the 2024 observations, we get marginal detections in individual lines with narrower wings in the 2023 spectra. Thus, it is evident that individual lines exhibit the same characteristics as suggested by the variation of $B_{\ell}$ in Figs. \ref{fig:GLP} and \ref{fig:all_Bz}. The intensity spectra of YZ Cet do not change significantly with time in individual lines. In Fig. \ref{fig:4b}, we show the variation of the LSD Stokes I profiles of the 2023 and 2024 observations and their residuals with respect to the median spectra. We do not notice any significant shift in radial velocity within the uncertainty (Fig. \ref{fig:4c}, bottom). However, we notice a rotational modulation of effective width (EW) in both datasets (Fig. \ref{fig:4c}, top), which could be a hint of an evolution also at the level of small-scale field.

\begin{figure*}
     \centering
     \begin{subfigure}[b]{0.48\textwidth}
         \centering
         \includegraphics[width=\textwidth]{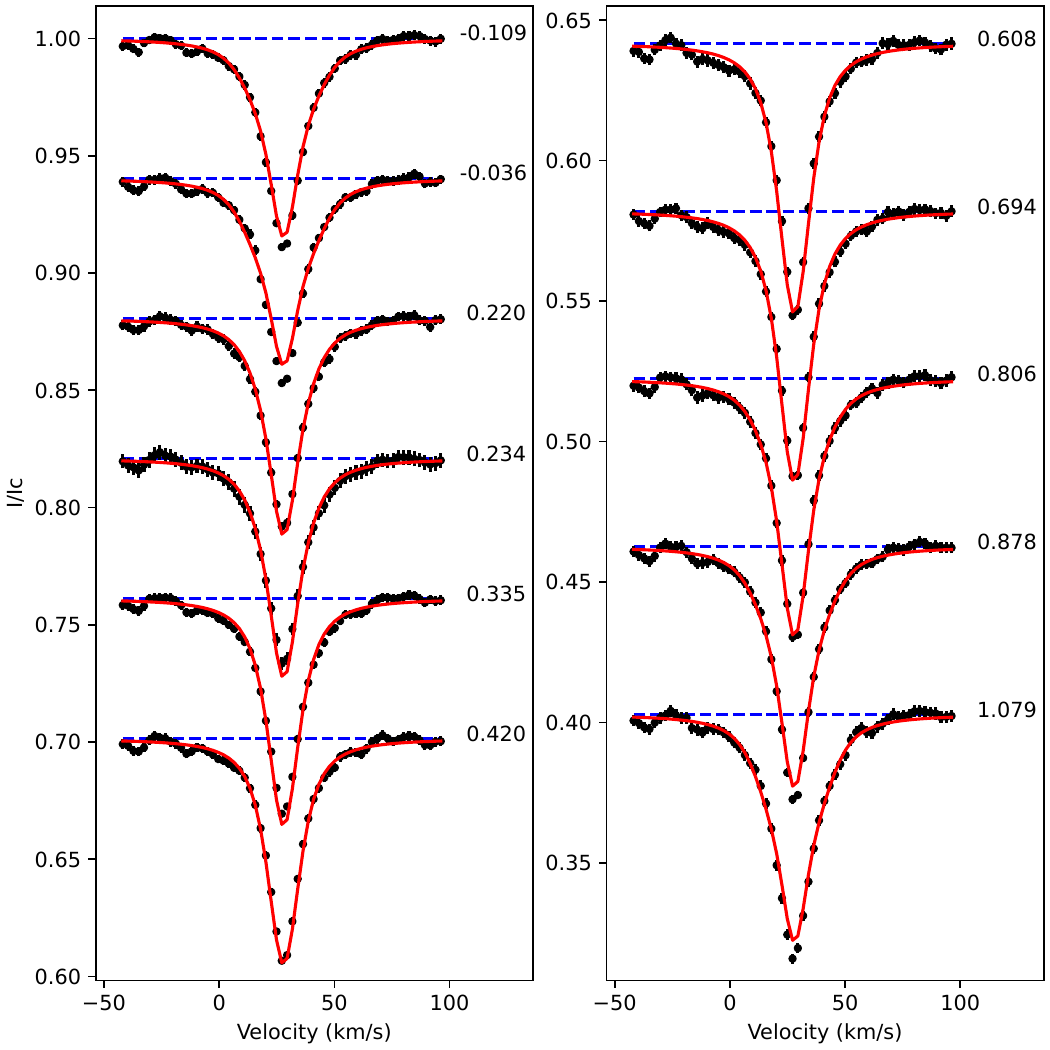}
         \caption{Stokes I ZDI model fit}
         \label{fig:ZDI_stokes_ABI}
     \end{subfigure}
     \begin{subfigure}[b]{0.48\textwidth}
         \centering
         \includegraphics[width=\textwidth]{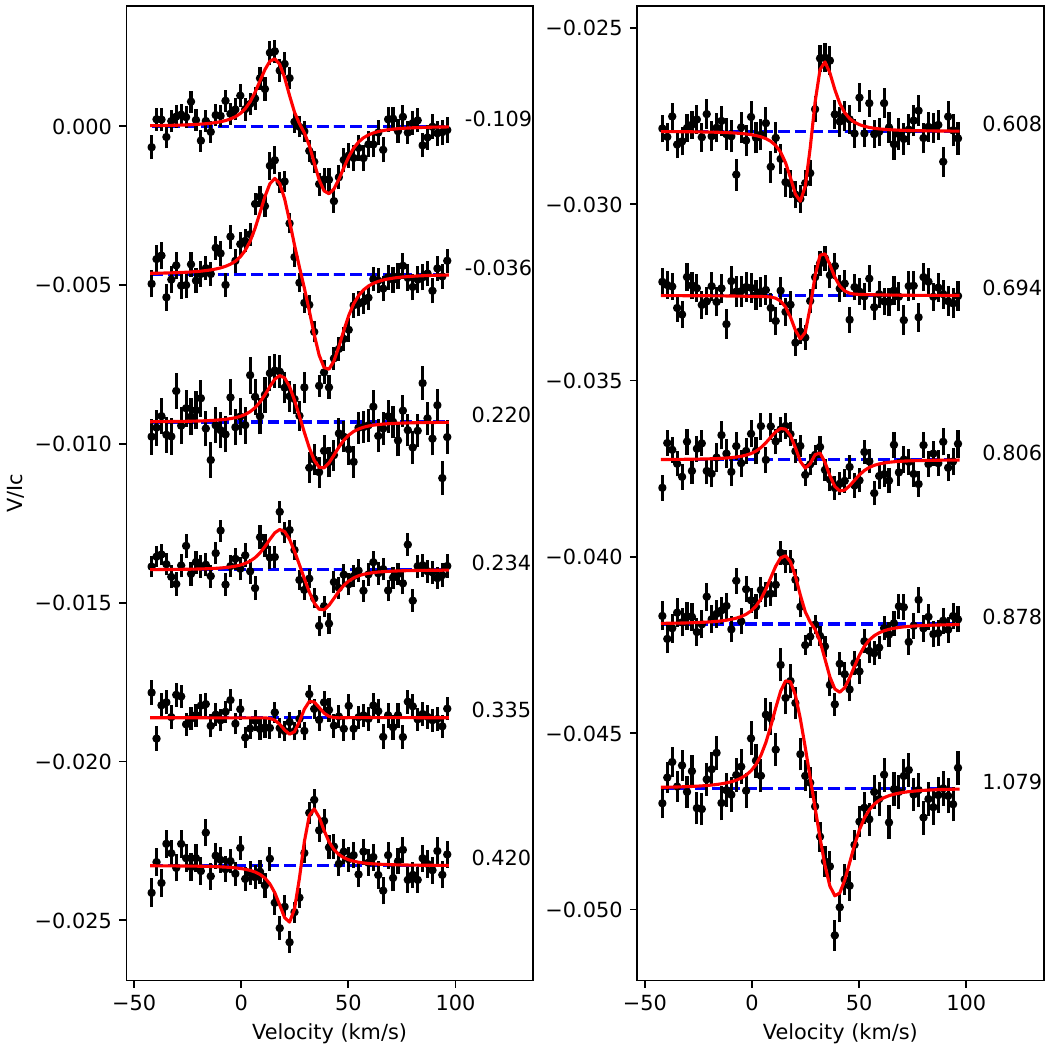}
         \caption{Stokes V ZDI model fit}
         \label{fig:ZDI_stokes_ABV}
     \end{subfigure}
        \caption{ZDI model fits (red curve) to the Stokes I (a) and Stokes V (b) LSD profiles (in black) from the 2024 SPIRou observations of YZ~Cet. Rotation cycles for each observation are labeled on the right of respective profiles. In both cases, the profiles are shifted vertically for clarity. \label{fig:ZDI_stokes_AB}}
\end{figure*}

\section{Magnetic Imaging} \label{ZDI}

\subsection{Unno–Rachkovsky Line Model}

We carried out a ZDI reconstruction \citep{Donati1997, Piskunov2002, Donati2006} of the surface magnetic field of YZ~Cet using two epochs of SPIRou observations to obtain further insight into the field topology and to quantitatively compare the change in global magnetic structure. The algorithm models the field as the sum of poloidal and toroidal components expressed through a spherical harmonic decomposition. Model Stokes V line profiles are generated, and then iteratively compared with observation to fit the spherical harmonic coefficients $\alpha_{\ell,m}$, $\beta_{\ell,m}$, and $\gamma_{\ell,m}$. The $\alpha_{\ell,m}$ coefficient corresponds to the radial poloidal field, $\beta_{\ell,m}$ the tangential poloidal field, and $\gamma_{\ell,m}$ the toroidal field ($\ell$ and $m$ being the degree and order of the mode, respectively). These parameters are fit until a target $\chi^2$ is reached, followed by an application of the maximum entropy regularization scheme to obtain the field map compatible with the lowest information content \citep{SkillingBryan1984, Donati1997, Folsom2018a}. For our analysis, we used a modified version of the Python-based program \texttt{ZDIpy} \citep{Folsom2018a}.

\begin{figure*}
     \centering
     \begin{subfigure}[b]{0.49\textwidth}
         \centering
         \includegraphics[width=\textwidth]{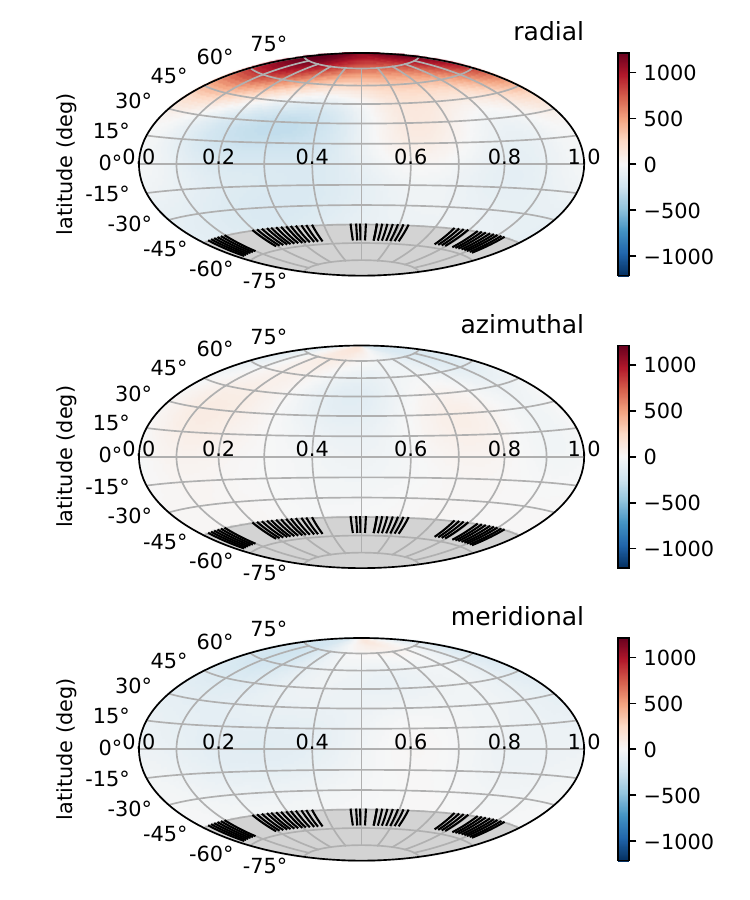}
         \caption{2023 SPIRou observations}
         \label{fig:ZDI_mag_oth}
     \end{subfigure}
     \begin{subfigure}[b]{0.49\textwidth}
         \centering
         \includegraphics[width=\textwidth]{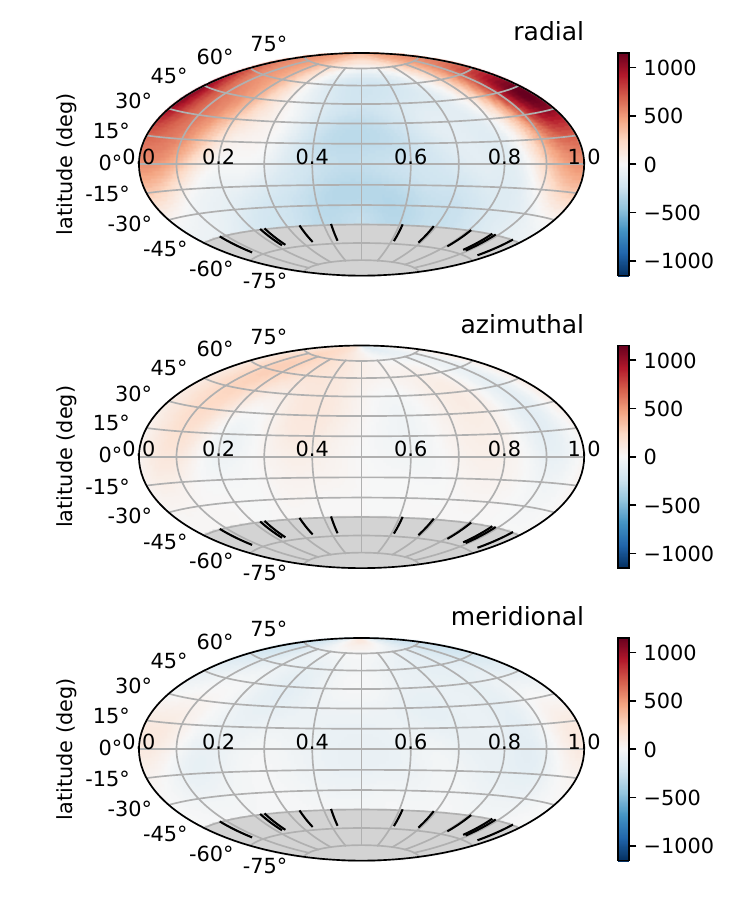}
         \caption{2024 SPIRou observations}
         \label{fig:ZDI_mag_AB}
     \end{subfigure}
        \caption{Magnetic map from ZDI modeling of SPIRou data of YZ~Cet presented in a Hammer projection. The top, middle and bottom panel represent the radial, azimuthal, and meridional components of the magnetic field. Tick marks at the bottom of the figures indicate the rotational phases at which observations were obtained. The gray area is the unobservable latitude of the star. \label{fig:ZDI_mag}}
\end{figure*}

For YZ~Cet, the widths of the Stokes V LSD profiles change with the observing phase and the longitudinal field strength. The profiles are broadest near phase 0.0, and narrowest near phase 0.5, indicating that a weak field model (such as that of \citealt{Folsom2018a}) is insufficient to reproduce the line broadening in the case of YZ Cet. In the modified \texttt{ZDIpy} code, we incorporate an Unno–Rachkovsky solution \citep{Unno1956, Rachkovsky1967} to the polarized radiative transfer equations in a Milne-Eddington atmosphere \citep{LandiLandolfi2004}, following the procedure described by \cite{Bellotti2023} and \cite{Erba2024}. The local line profiles in this model are described by the Lorentzian width ($w_L$), related to pressure broadening, the Gaussian width ($w_G$) which is related to thermal broadening, the ratio of the line to the continuum absorption coefficients ($\eta_0$), and finally the slope of the source function in the Milne–Eddington atmosphere, $\beta$ \citep{LandiLandolfi2004}. As we are fitting LSD profiles instead of actual spectra, the fiducial line is assumed to split as a Zeeman triplet.

The model treats the limb darkening of the continuum and the variation in line strength across the stellar disk as separate effects, both described using the Milne-Eddington approximation. The continuum intensity at an angle $\theta$ from the disk center is given by

\begin{equation}
    I_c = B_0(1 + \beta \cos \theta),
\end{equation}

\noindent
where $B_0$ is the surface value of the source function \citep{LandiLandolfi2004}. Normalized to the intensity at disk center, this becomes

\begin{equation}
    \frac{I_c}{I_c^0} = \frac{1 + \beta \cos \theta}{1 + \beta}.
\end{equation}

To be consistent with standard limb darkening laws, we compared this with the linear limb darkening prescription,

\begin{equation}
    \frac{I_c}{I_c^0} = 1 - \eta + \eta \cos \theta,
\end{equation}

\noindent
where $\eta$ is the limb-darkening coefficient. Equating the two expressions, we obtain $\beta = \frac{\eta}{1 - \eta}$. This allows us to set $\beta$ based on tabulated $\eta$ values. For all ZDI reconstructions, we adopted $\eta = 0.3$ in the H band \citep{ClaretBloemen2011}.

We also adopt the filling factor formalism adopted by \cite{Morin2008} to improve the fit to the observed Stokes I and Stokes V profiles. In this formalism, two additional parameters $f_I$ and $f_V$ are included, where $f_I$ represents the fraction of the surface being magnetic, and $f_V$ is the fraction of the surface that contributes to the net circular polarization signature.

\begin{figure*}
    \centering
    \includegraphics[width=0.95\linewidth]{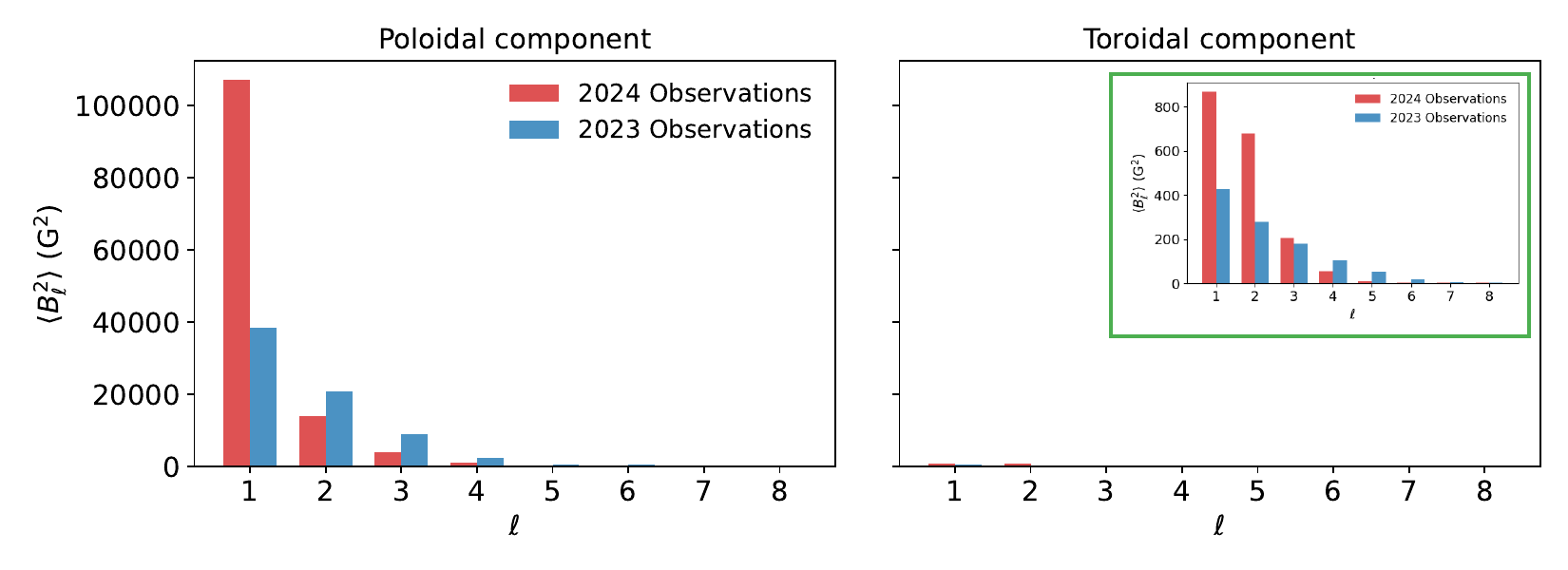}
    \caption{Comparison of the magnetic energies by degree $\ell$ for magnetic maps for 2023 (blue) and 2024 (red) SPIRou observations separated into poloidal (left) and toroidal (right) components. In the right plot, we show a zoomed view in the inset. { The plotted quantity $\langle B^2 \rangle = \oint \mathbf{B} \cdot \mathbf{B} \, d\Omega/4\pi$ in each spherical harmonic of degree $\ell$ is proportional to the true magnetic energy.}  }
    \label{fig:map_compare}
\end{figure*}

\subsection{Optimizing Parameters}

For this analysis, we adopted the primary ephemeris obtained from the GRP fit, as discussed in Section \ref{Bz_vary}. The mean RV values were adopted during the fit (see Fig. \ref{fig:4c}).  We used the same average central wavelength $\lambda_0 = 1200$ nm and an effective Land\'e factor $g_{\rm eff} = 1.2$ as in the LSD analysis. { We restrict the degree of spherical harmonics of the ZDI reconstructions to $\ell = 8$ for flexibility, although harmonics above $\ell = 3$ or 4 are likely unresolved.  We allow the maximum entropy regularization to send poorly constrained harmonics to near-zero values.} We do not simultaneously fit Stokes I, and we assume a uniform brightness. This is motivated by the fact that Stokes I does not vary significantly over rotation phase (Fig. \ref{fig:4b}). { However, we re-normalized the Stokes I profiles to remove the pseudo-continuum level, which can be attributed to the contribution of weak molecular lines.} 

Since observations only provide an upper limit on $v \sin{i}$ ($<2$ km/s, \citealt{Reiners2018}), we adopted the equatorial velocity of the star ($\sim 0.11$ km/s) calculated using the radius ($R_* = 0.157 \pm 0.005 R_{\odot}$) derived by \cite{Schweitzer2019} and the period derived in this study as the initial proxy of $v \sin{i}$. This low value of $v \sin{i}$ also prevented us from reliably deriving the inclination ($i$). Thus, we start with $i=65^{\circ}$, motivated by the large variation of $B_{\ell}$. 

For each observation epoch, we optimize various parameters (e.g. $w_L, \ w_G,$ and $\eta_0$) by minimizing the $\chi^2$ difference between the synthetic and observed Stokes~$I$ profiles. A grid of synthetic profiles was generated and compared with the median observed profile, and the set of parameters that produced the lowest $\chi^2$ was adopted for the ZDI reconstruction. The best-fit Gaussian and Lorentzian widths were $w_G = 2.1$ km/s and $w_L = 6.25$ km/s, respectively. The inclination was restricted using the value that optimizes the ZDI results of both epochs following \cite{Folsom2018b, Folsom2020}. This was performed by running fits correspond to a grid of inclinations by (a) first converging to a target $\chi^2_r$ (set to 1.1), and (b) then for a fixed best-fit entropy, we repeated the process using a smaller grid, and fit the inclination by minimizing $\chi^2_r$. {  Based on this fitting, we find $i=50 \pm 10^{\circ}$, which we used for the final ZDI map. We also fit the Stokes V best-fit filling factor ($f_V$) using a similar process. For the 2024 dataset, we found $f_V=11 \pm 1$\%. However, we did not find a well-defined optimal solution (i.e. a minimum in $\chi^2$ while fixing the entropy) for $f_V$ in the 2023 data. In this case, a smaller $f_V$ is always favored. We therefore used a value $f_V= 14$\% that is closest to the 2024 epoch value, for which we get a converging solution. }

\begin{deluxetable}{lcc}
\tablecaption{Parameters describing the magnetic map.  \label{tab:ZDI}}
\tablehead{
\colhead{Magnetic Parameter} & \colhead{2023} & \colhead{2024} 
} 
\startdata 
Mean $|B|$ & 219 G &  298 G  \\
{ Mean radial $|B_r|$} & { 201 G} &  { 281 G}  \\
Maximum $|B|$ & 859 G & 834 G \\
Dipole $B_p$ & 280 G & 430 G \\
Dipole obliquity & $17.5^{\circ}$ & $51.5^{\circ}$  \\
Dipolar energy & 53\% & 75\%  \\
Quadrupolar energy & 29\% & 19\% \\
$f_I$ & 50\% & 50\%  \\
$f_V$ & $<14$\% & $11 \pm 1$\%  \\
\multicolumn{3}{c}{Distribution of magnetic energy} \\
Poloidal (\% tot) & 99\% & 99\%   \\
Toroidal  (\% tot) & 1\% & 1\%   \\
Axisymmetric (\% tot) &  87\% & 34\% \\
\enddata
\end{deluxetable}

\subsection{Best Fit Maps}

The best fit to the Stokes I and Stokes V LSD profiles for the 2024 epoch is shown in Fig. \ref{fig:ZDI_stokes_AB}, while the same for the 2023 epoch is given in Appendix \ref{App_C}. The corresponding best-fit magnetic maps for the 2023 and 2024 SPIRou epochs are shown in Fig. \ref{fig:ZDI_mag}. The magnetic geometry is described in terms of ratios of magnetic energy density $\langle B^2 \rangle = \oint \mathbf{B} \cdot \mathbf{B} \, d\Omega/4\pi$ (Fig. \ref{fig:map_compare}). The parameters describing the magnetic map are presented in Table~\ref{tab:ZDI}. { The magnetic field values are effectively spatially averaged, analogous to a magnetic flux density (i.e. locally magnetic fields reach $B/f_V$, for the reported $B$ and filling factor $f_V$)}. In both epochs, the magnetic map is found to be predominantly poloidal (poloidal magnetic energy $\sim 99$\%). The toroidal contribution is $\sim 1$\% in both epochs. The extremely low $v \sin{i}$ of the star makes the toroidal field nearly undetectable. Thus the very low toroidal fraction may only reflect the detection limits of our observations, and not the real absence of toroidal field in the star.

Varying the inclination by $\pm 10$ degrees and calculating magnetic maps leads to an uncertainties in $B_{\rm mean}$ of $\sim40$ G, $B_{\rm max} \sim 50$ G, and $B_{\rm pol} \sim 90$ G,  with uncertainties in total axisymmetry of $\sim4$\% and in the poloidal fraction $<1$\%. Varying $f_V$ by $\pm 1$\%  and calculating magnetic maps leads to uncertainties in $B_{\rm mean}$ of $\sim 6$ G, $B_{\rm max} \sim 40$ G, and $B_{\rm pol} \sim 15$ G, with uncertainties in total axisymmetry of $\sim 3$\% and in the poloidal fraction $<1$\%.

\begin{figure}
    \centering
    \includegraphics[width=0.98\linewidth]{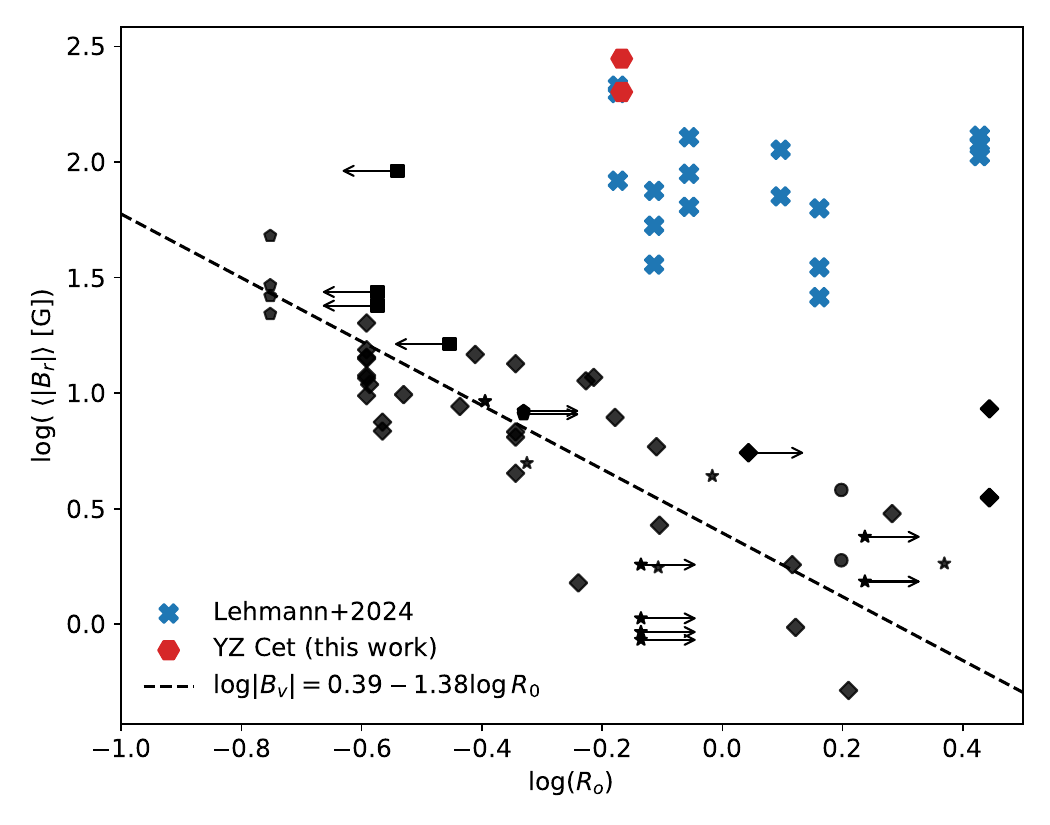}
    \caption{Location of YZ Cet among the correlation diagram between the average large-scale radial field strength ($\langle B_r \rangle$) and Rossby number $R_o$ by \cite{Vidotto2014}. The figure includes a sample of slowly rotating M dwarfs (in blue) reported in \cite{Lehmann2024}. The mean field during the 2024 epoch of our observations place YZ Cet as the host of the strongest field. For details on the symbol of black markers, see Fig. 4 of \cite{Vidotto2014}. { For this figure, $R_o$ was determined using the formulation by \cite{Wright2011} for consistency with \cite{Lehmann2024} and \cite{Vidotto2014}. For the rest of the paper, we used the latest formalism of \cite{Wright2018}.}  }
    \label{fig:Vidotto}
\end{figure}

We noticed a significant change in the overall magnetic strength and geometry of the two epochs (Fig. \ref{fig:ZDI_mag}). The average surface magnetic field strength changes from $\sim 219$ G in the 2023 data to $\sim298$ G in the 2024 observations. The dipolar ($\ell = 1$) energy contribution changed from $\sim53$\% to $\sim75$\%, while the dipole polar field strength also increased from $\sim 280$ G in the 2023 data to $\sim 430$ G in the 2024 observations. We find that YZ~Cet hosts the strongest dipolar field among the known slowly rotating M dwarf systems (Fig. \ref{fig:Vidotto}). During the 2023 epoch, the field is largely axisymmetric about the rotation axis, with $\sim 87$\% of the energy in the $m=0$ components. The axisymmetry drastically changed, with $<35$\% of the magnetic energy in the $m=0$ component in the 2024 observations. This dramatic change is reflected in dipolar obliquity (i.e., the angle between the magnetic and rotational axes), which changed from $17.5^{\circ}$ to $51.5^{\circ}$ in less than a year, suggesting the possibility of a future/imminent polarity reversal. During the 2023 epoch, we observe a significant quadrupole ($\ell= 2$) component, containing $\sim30$\% of the total poloidal energy. In the 2024 observations, this decreases to $\sim 19$\%, making the field predominantly dipolar. In both cases, the maximum surface field strength is found to be similar.

\section{Constraint on Planetary Field}

\cite{Trigilio2023} observed YZ~Cet in low-frequency radio bands with uGMRT (550–900 MHz range) at nine epochs and detected radio bursts four times. Two of these bursts exhibited high circular polarization and, when phase-folded on the orbital period of YZ~Cet~b ($P_{\rm orb} = 2.02\,$days), were clustered in two narrow orbital sectors (phases 0.07–0.20 and 0.78–0.90), consistent with expectations from a hollow-cone ECME model. Under the assumption of sub-Alfvénic SPI, the authors modeled the emission with a dipolar stellar field ($B_p \approx 2.4\,$kG). This first estimate of the dipolar field strength was only an approximate value, as it relied on the assumption of the radius of origin and cut-off frequencies. The observed ECME power was found to be $P_{\rm obs} \approx 1.0$--$1.7\times10^{22}$ erg~s$^{-1}$. Assuming an incident power $P_{\rm in}$ due to the interaction between YZ~Cet's magnetosphere and the planet YZ~Cet~b, we can estimate the radius of the planet { \citep{Trigilio2023}}:

\begin{equation}
    R = \sqrt{\frac{8 P_{\rm in}}{v_{\rm rel} B^2}},
\end{equation}

\noindent
where $v_{\rm rel}=85.1$ km/s is the velocity of the planet relative to the stellar magnetic field, and $B$ is the approximate stellar field strength around the planet. To satisfy $P_{\rm in} > P_{\rm obs}$, \cite{Trigilio2023} proposed that a magnetic shielding mechanism must be present. This can effectively enhance the cross section of the planetary body to a magnetosphere of the planet, thus requiring a magnetopause radius ($R_{\rm MP}$) in place of $R$. Assuming a dipolar magnetic field for both the star and the planet, we can then estimate the planetary polar magnetic field strength { based on  \cite{Trigilio2023}} as:

\begin{figure}
    \centering
    \includegraphics[width=0.94\linewidth]{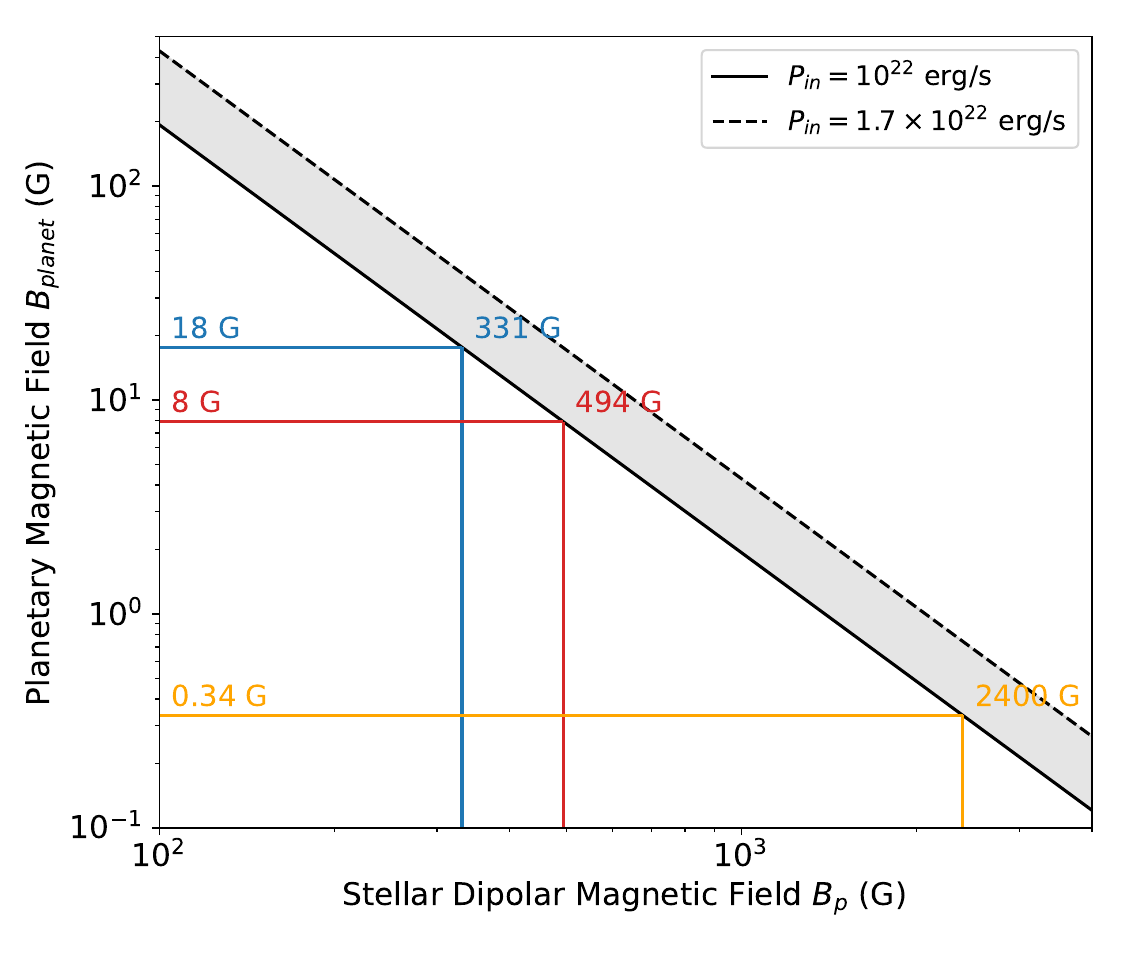}
    \caption{Minimum planetary magnetic field ($B_{\rm planet}$) required to power the observed radio emission from YZ~Cet~b, as a function of stellar dipolar magnetic field ($B_p$). The two curves represent different power estimates ($P_{\rm in}= 10^{22}$ and $1.7 \times 10^{22}$  erg/s), with the shaded region indicating the allowed range. Vertical dashed lines show specific Bp values (331 G, 494 G, 2400 G, corresponding to the 2023, 2024 dataset, and indirect measurement from radio observations), and horizontal lines mark the corresponding $B_{\rm planet}$ lower limits.}
    \label{fig:planet_field}
\end{figure}

\begin{equation}
    B_{\rm planet} = \frac{32}{R_{\rm planet}^3 B_p^2} \cdot \left( \frac{r}{R_*} \right)^6 \cdot \left( \frac{2 P_{\rm in}}{v_{\rm rel}} \right)^{3/2},
\end{equation}

\noindent
where $R_{\rm planet}$ is the radius of YZ~Cet~b, $R_*$ is the stellar radius. $r$ is the average distance between the star and the planet, and $B_p$ is the dipolar field strength of the star. In Fig. \ref{fig:planet_field}, we show the predicted planetary field strength given different $B_p$ values. We estimate lower limits for the planetary field to be 18 G and 8 G from the 2023 and 2024 observations, respectively. This estimate is much higher than the predicted value of 0.4 G by \cite{Trigilio2023}.

\begin{figure*}
    \centering
    \includegraphics[width=0.8\linewidth]{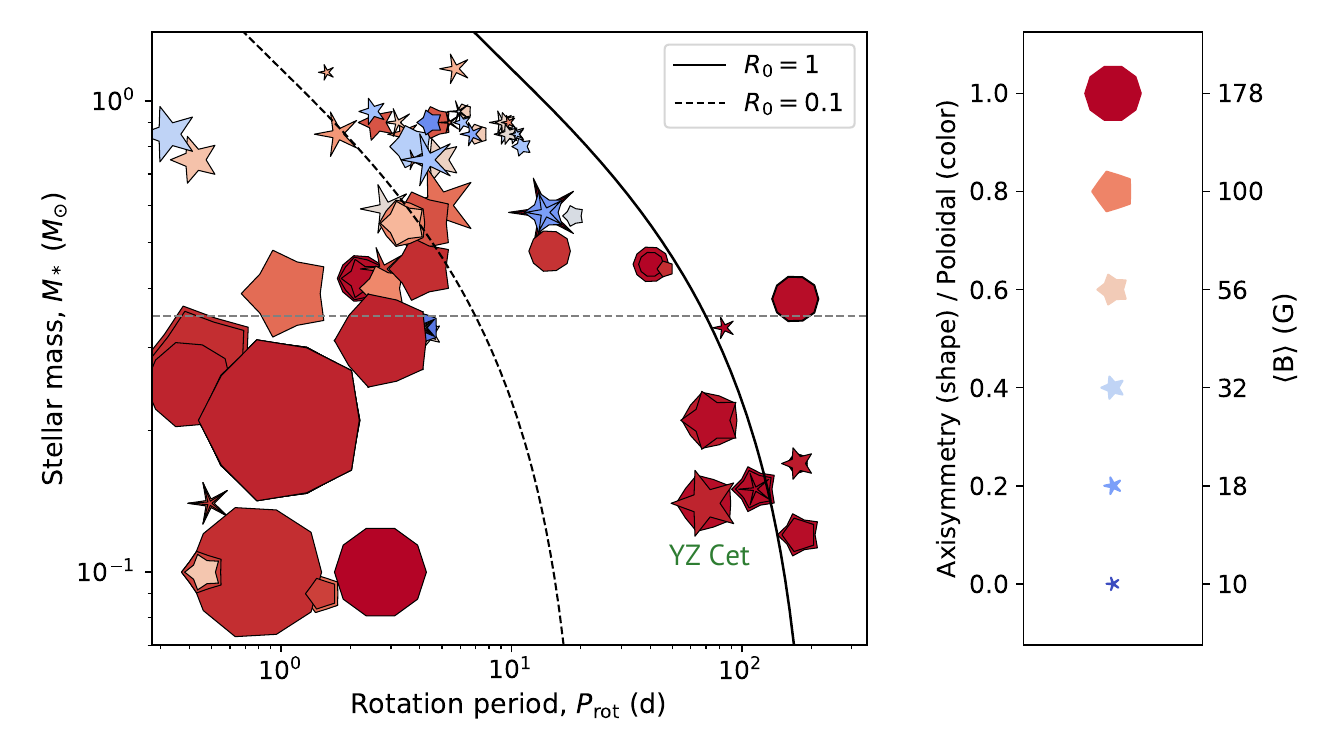}
    \caption{Location of YZ Cet among other M-dwarfs as a function of stellar mass and rotation period. The size of each symbol represents the strength of the large-scale magnetic field, while its shape indicates the level of axisymmetry, with decagons corresponding to fully axisymmetric fields and star shapes to completely non-axisymmetric ones. The symbol’s color provides an approximate indication of the magnetic field’s topology, ranging from red (purely poloidal) to blue (purely toroidal). The black solid line and dashed line show where the Rossby number equals 1 and 0.1, respectively \citep{Wright2018}. The gray dashed line indicates the $\sim0.35\,M_{\odot}$ mass threshold below which M dwarfs are fully convective \citep{Chabrier1997}. Other data points used in these plots are taken from { \cite{Donati2008b, Morin2008, Morin2010, Morin2011, Folsom2016, Folsom2018a, Lehmann2024} and references therein}.}
    \label{fig:YZ_cet_confusogram1}
\end{figure*}

As the magnetic field is rapidly evolving with time, it is not possible to reliably predict the planetary field without simultaneous radio and spectropolarimetric observations. The lower limit predictions heavily depend on the stellar field strength and observed radio luminosity (Fig. \ref{fig:planet_field}). Additionally, it should be noted that all calculations are based on the assumption that both stellar and planetary fields are roughly aligned and dipolar. In reality, this assumption is not generally valid; however, since the strengths of the higher-order components of the field decrease rapidly along the radial direction, the contribution from the dipolar component likely dominates at the distance of the planet. For a more realistic estimation, it is necessary to take into account the topology of the magnetic field of both the star and the planet (e.g. \citealt{Strugarek2015, Strugarek2022, Vidotto2025}).

\section{Discussion and Conclusions} \label{Discussion}

Our spectropolarimetric monitoring and magnetic imaging of YZ~Cet have revealed substantial evolution in its large-scale magnetic topology over a relatively short timescale. We observe a clear short-term evolution in $B_{\ell}$ over a timescale of a few rotation cycles, with the peak-to-peak amplitude of the $B_{\ell}$ measurements increasing from $\sim200$ G in 2023 to $\sim400$ G in 2024. The magnetic field evolution is further confirmed by the ZDI maps. We observe a significant increase in the polar magnetic field strength, from $\sim$332 G to $\sim$494 G and in the average surface magnetic field strength from $\sim$201 G to $\sim$276 G between the 2023 and 2024 epochs. The topology remains predominantly poloidal in both epochs (with $\geq$96\% of the magnetic energy in the poloidal modes), but the geometry and axisymmetry of the field undergo substantial changes. The dipolar energy fraction increases from $\sim$54\% to $\sim$77\%, while the contribution of the quadrupolar component ($\ell=2$) decreases, indicating a simplification of the field structure toward a more ordered dipolar configuration. On the other hand, the axisymmetry drops from $\sim$83\% to $<35$\%. Such rapid changes in the orientation and symmetry of the magnetic field suggest a possible ongoing polarity reversal, although continued monitoring will be required to determine whether the global magnetic field varies cyclically or evolves more chaotically.

Previous ZDI studies have revealed diverse magnetic topologies in both partly and fully convective M dwarfs \citep{Donati2008, Donati2008b, Morin2008, Morin2010, Hebrard2016}. For fully convective stars with $M < 0.2 \ M_{\odot}$, these studies have identified a striking dichotomy: some stars possess strong, largely axisymmetric, dipole-dominated fields, while others display weaker, non-axisymmetric fields with complex multipolar structures \citep{Morin2008, Morin2010}. This pattern has been interpreted in two main ways. One interpretation is dynamo bistability, in which two distinct and stable large-scale magnetic configurations coexist in parameter space \citep{Morin2011, Gastine2013, Kochukhov2017}. Another possibility is that stars undergo long-term magnetic cycles, including global polarity reversals, and that observations sometimes capture them at different phases of such a cycle \citep{Kitchatinov2014}. Although the large-scale magnetic field topology and evolution of rapidly rotating ($R_{O}\lesssim0.1$) M dwarfs has been reasonably well studied, only a few very slow rotators ($R_{O}\sim1$) have been mapped with ZDI over multiple rotational cycles. 

YZ~Cet does not fall cleanly into either branch of the proposed bistability. In its weaker-field state, the field is not as weak or as complex as the multipolar branch, whereas in its stronger-field state it is not as purely axisymmetric as the dipolar branch. In addition, the magnetic topology of YZ~Cet appears to change more rapidly than in most stars associated with bistability, although additional epochs will be needed to confirm the characteristic variability timescale. The rotation rate is likely to be an important factor in this case. YZ~Cet rotates much more slowly than the rapidly rotating fully convective M dwarfs that motivated the bistability hypothesis. If rapid rotation is essential for generating a stable dichotomy of states, then stars with slower rotation may instead produce magnetic fields that are less stable in time, evolving more rapidly and perhaps more chaotically. This interpretation is consistent with the models of \citet{Gastine2013}, which find bistability primarily at smaller Rossby numbers (roughly below 0.1). At larger Rossby numbers, their simulations produce more multipolar fields, a configuration that resembles the geometry we observed in YZ~Cet in 2023. However, these models do not fully reproduce the strong, non-axisymmetric field observed in 2024, suggesting that additional processes may be involved. 

Fig. \ref{fig:YZ_cet_confusogram1} places YZ~Cet in the broader context of M dwarfs with measured large-scale magnetic fields. Its position far from the fully convective boundary in mass, combined with its relatively long rotation period, places it well outside the region where bistability has been observed in rapid rotators. The figure also highlights that YZ~Cet hosts one of the strongest large-scale fields among slowly rotating low-mass M dwarfs, while also exhibiting substantial deviations from axisymmetry in some epochs. Fig. \ref{fig:Vidotto} also highlights the position of YZ~Cet relative to the empirical trend from \citet{Vidotto2014} that suggests that its magnetic field is atypically strong for its rotation rate. This supports the interpretation that YZ~Cet may represent a different dynamo operating regime from that of both rapidly rotating fully convective stars and slowly rotating partly convective stars.  

The slow rotation and relatively rapid magnetic variability of YZ~Cet make it more comparable to the stars studied by \citet{Lehmann2024} than to those in the bistability regime. In the Lehmann et al. sample, three slowly rotating fully convective M dwarfs exhibit stable, predominantly axisymmetric large-scale fields over timescales of at least one year, while another three display more variable magnetic fields. One of the variable stars, GJ~1151, shows a possible polarity reversal. YZ~Cet shares the property of rapid magnetic variability with this second group, but its magnetic field is significantly stronger. For example, the dipolar field strength in GJ~1151 ranges from $23$ to $62$~G, whereas YZ~Cet reaches several hundred gauss. YZ~Cet also likely has a smaller Rossby number than any of the Lehmann et al. stars. This could contribute to its stronger field, although among slowly rotating fully convective M dwarfs the relationship between Rossby number and magnetic field strength appears weak.  

Taken together, these similarities and differences suggest that YZ~Cet and the variable stars studied by \citet{Lehmann2024} may occupy a distinct dynamo regime, separate from both rapidly rotating fully convective stars that exhibit bistability and slowly rotating partly convective stars. The magnetic fields in this regime appear capable of evolving significantly on timescales of months to years, but it is not yet clear whether the changes follow a cyclic pattern or occur in a more stochastic manner. Long-term spectropolarimetric monitoring of YZ~Cet and similar stars will be crucial for constraining the physical mechanisms operating in this regime and for testing whether their dynamo behavior is governed by principles different from those of other fully convective or partly convective M dwarfs.

We were able to place a more realistic lower bound on the planetary magnetic field strength by combing direct estimates of stellar dipolar field strengths from spectropolarimetric observations with existing radio observations. While \citet{Trigilio2023} suggested a planetary field of approximately 0.4\,G, our analysis yields revised lower limits of $\sim$18\,G and $\sim$8\,G, corresponding to the 2023 and 2024 magnetic field measurements, respectively. However, all such estimates assume idealized dipolar topologies for both the star and planet. Given that real magnetic field configurations are more complex, future modeling must account for higher-order components and non-axisymmetries. Also, given the evolving nature of the stellar magnetic field, simultaneous spectropolarimetric and radio monitoring will be essential to reliably constrain the planetary magnetosphere and to better understand the magnetic interactions in this system. Recently, \cite{Pineda2025} has independently analyzed the 2023 data, and their maps are consistent with this study. The authors also show that the radio emission is roughly consistent with ECME from SPI.

\begin{acknowledgments}
CPF acknowledges funding from the European Union’s Horizon Europe research and innovation programme under grant agreement No. 101079231 (EXOHOST), and from the United Kingdom Research and Innovation (UKRI) Horizon Europe Guarantee Scheme (grant number 10051045). JAB acknowledges support through an NSERC Postgraduate Doctoral Scholarship (PGS D). GAW acknowledges support in the form of a Discovery Grant from the Natural Sciences and Engineering Research Council (NSERC) of Canada.  SB acknowledges funding by the Dutch Research Council (NWO) under the project "Exo-space weather and contemporaneous signatures of star-planet interactions" (with project number OCENW.M.22.215 of the research programme "Open Competition Domain Science- M").

All data exploited in this paper are available from the corresponding public archives. This work has made use of the VALD data base operated at Uppsala University, the Institute of Astronomy RAS in Moscow, and the University of Vienna. This research has made use of NASA’s Astrophysics Data System Bibliographic Services, and of the SIMBAD database, operated at CDS, Strasbourg, France. This work is based on observations obtained with the Canada–France–Hawaii Telescope (CFHT), which is operated by the National Research Council (NRC) of Canada, the Institut National des Sciences de l’Univers of the Centre National de la Recherche Scientifique (CNRS) of France, and the University of Hawaii. The observations at the CFHT were performed with care and respect from the summit of Maunakea, which is a significant cultural and historic site.
\end{acknowledgments}

\begin{contribution}

AB has conceptualized the project, wrote telescope proposals, led the data reduction and analysis, and primarily wrote the manuscript. CPF guided AB during writing telescope proposals and made the line list for LSD analysis. JAB analyzed archival ESPaDOnS data and help schedule observations. All authors shared ideas and contributed while writing the manuscript. 

\end{contribution}

\facilities{CFHT (ESPaDOnS, SPIRou), TBL (NARVAL)}

\software{astropy \citep{Astropy2022}, 
          numpy \citep{harris2020array}, 
          scipy \citep{Virtanen2020},
          emcee \citep{Foreman-Mackey2013}
          SpecpolFlow \citep{Folsom2025}
          }

\bibliography{sample7}{}
\bibliographystyle{aasjournal}

\appendix

\section{ZDI Line Fits to the 2023 Observations} \label{App_C}

\begin{figure*}
     \centering
     \begin{subfigure}[b]{0.45\textwidth}
         \centering
         \includegraphics[width=\textwidth]{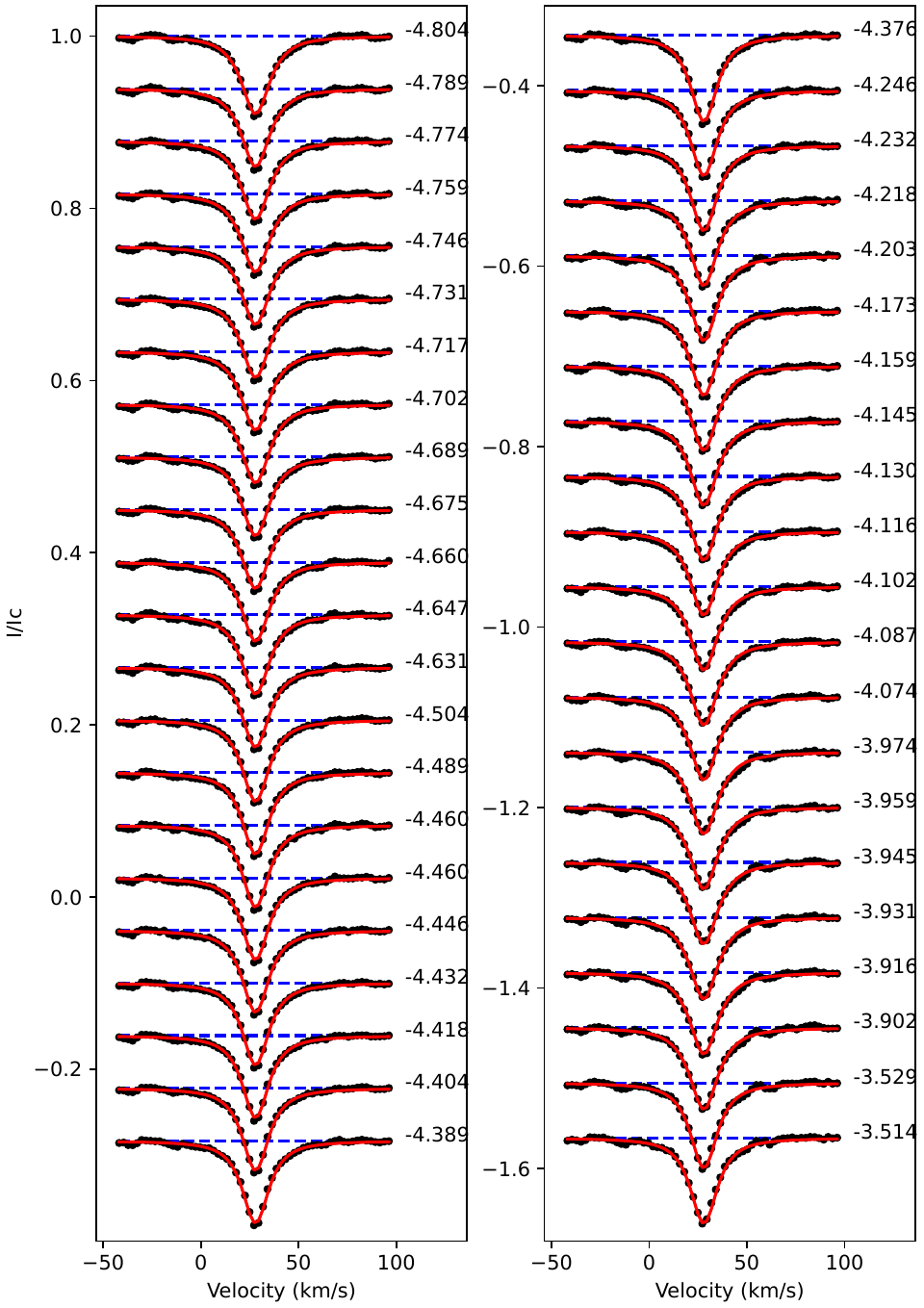}
         \caption{Stokes I ZDI model fit (2023)}
         \label{fig:ZDI_stokes_othI}
     \end{subfigure}
     \begin{subfigure}[b]{0.45\textwidth}
         \centering
         \includegraphics[width=\textwidth]{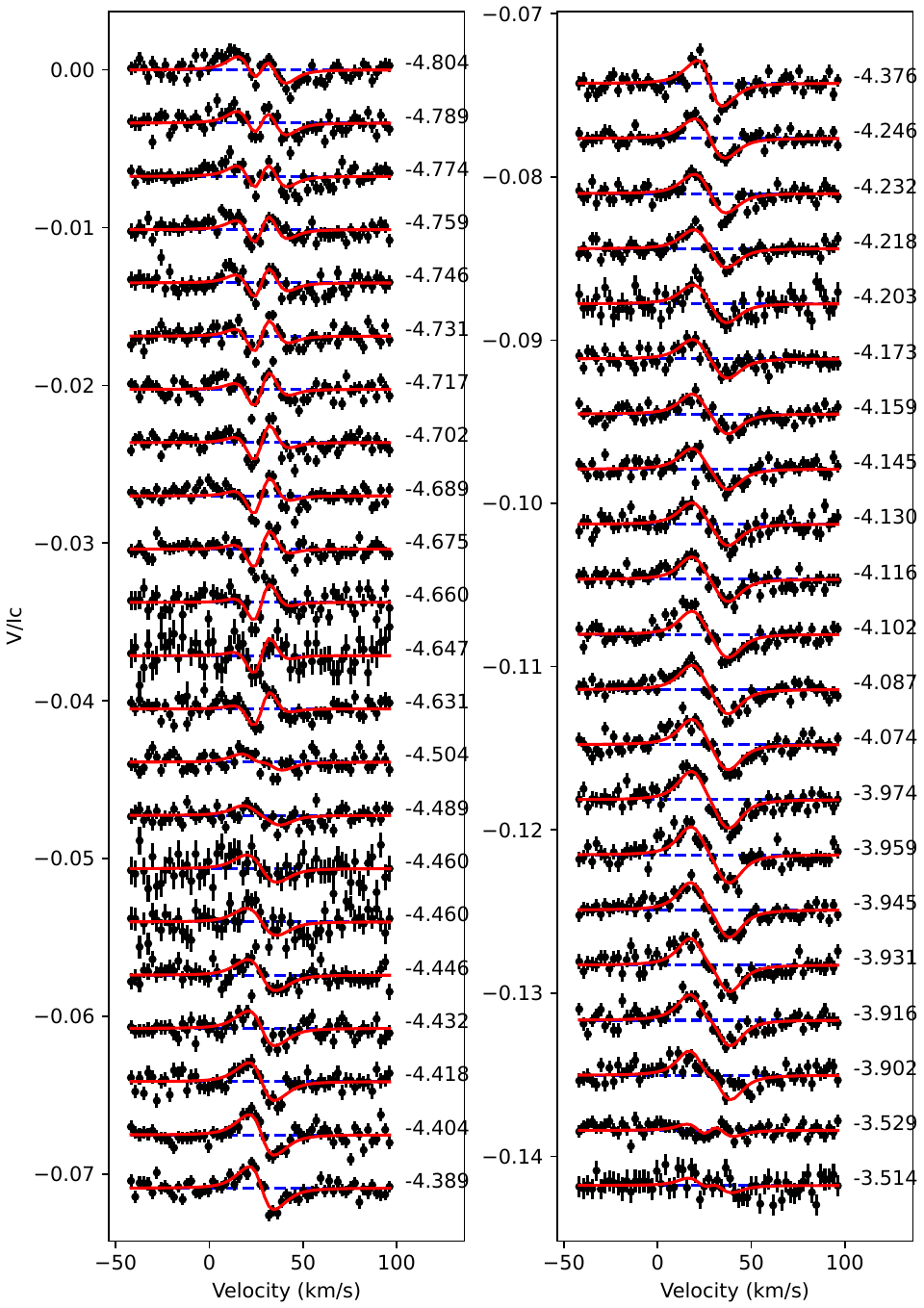}
         \caption{Stokes V ZDI model fit (2023)}
         \label{fig:ZDI_stokes_othV}
     \end{subfigure}
        \caption{Same as Fig. \ref{fig:ZDI_stokes_AB} for the 2023 SPIRou observations.} \label{fig:ZDI_stokes_oth}
\end{figure*}

In Fig. \ref{fig:ZDI_stokes_oth}, we show the ZDI model fits to the Stokes I Stokes V LSD profiles from the 2023 SPIRou observations of YZ Cet.

\end{document}